\documentclass[aps,pra,twocolumn,showpacs,preprintnumbers,superscriptaddress]{revtex4-1}
\usepackage{amsmath}
\usepackage{epsfig}
\usepackage{graphicx}
\usepackage{color}
\usepackage{amssymb}
\usepackage{epstopdf}
\usepackage{array}
\usepackage{tabularx}
\usepackage[utf8]{inputenc}
\usepackage[T1]{fontenc}
\begin{document}
\title{Vortex generation in stirred binary Bose-Einstein condensates}
\author{Anac\'e N. da Silva}
\affiliation{
Instituto de F\'isica Te\'orica, Universidade Estadual Paulista, 01156-970 
S\~ao Paulo, SP, Brazil.}
\author{R. Kishor Kumar}
\affiliation{
Department of Physics, Centre for Quantum Science, and Dodd-Walls Centre for
Photonic and Quantum Technologies, University of Otago, Dunedin 9054, New Zealand.}
\author{Ashton S. Bradley}
\affiliation{
Department of Physics, Centre for Quantum Science, and Dodd-Walls Centre for
Photonic and Quantum Technologies, University of Otago, Dunedin 9054, New Zealand.}
\author{Lauro Tomio}
\affiliation{
Instituto de F\'isica Te\'orica, Universidade Estadual Paulista, 01156-970 
S\~ao Paulo, SP, Brazil.}
\date{\today}
\begin{abstract}
 The dynamical vortex production, with a trap-confining time-dependent stirred potential, is studied by using 
 mass-imbalanced cold-atom coupled Bose-Einstein condensates (BEC). The vortex formation is explored in the 
 condensate laboratory frame, by considering that both coupled species are confined by a pancake-like harmonic 
 trap, slightly modified elliptically by a time-dependent periodic potential, with the characteristic frequency enough 
 larger than the transversal trap frequency. The approach is applied to the experimentally accessible binary mixtures 
 $^{85}$Rb-$^{133}$Cs and $^{85}$Rb-$^{87}$Rb, which allow us to verify the effect of mass differences in the 
 dynamics. For both species, the time evolutions of the respective energy contributions, together with associated 
 velocities, are studied in order to distinguish turbulent from non-turbulent flows. By using the angular momentum 
 and moment of inertia mean values, effective classical rotation frequencies are suggested, which are further 
 considered within simulations in the rotating frame without the stirring potential.  A spectral analysis is also 
 provided for both species, with the main focus in the incompressible kinetic energies. In the transient turbulent 
 regime, before stable vortex patterns are produced, the characteristic $k^{-5/3}$ Kolmogorov 
 behavior is clearly identified for both species at intermediate momenta $k$ above the inverse Thomas-Fermi 
 radial positions, further modified by the universal $k^{-3}$ scaling at momenta higher than the inverse of the 
 respective healing lengths.  
 Emerging from the mass-imbalanced comparison, relevant is to observe that the verified change in the spectral 
 scaling behavior occurs at lower incompressible kinetic energies for larger mass differences between the species; 
 consequently, as larger is the mass difference, much faster is the dynamical production of stable vortices.
 \end{abstract}
\flushbottom
\maketitle
\section{Introduction}
\label{secI}
The experimental observations of vortices in Bose-Einstein condensates (BEC)~\cite{1999Matthews,2000MadisonPRL,2000Chevy,2000Anderson,2001Madison,2001Hodby,2001Shaeer,2006Bewley}, 
have been motivated by looking to their predicted relevance on fundamental aspects of quantum mechanics,
such as superfluidity, which was previously verified in rotating superfluid of helium, with the formation 
of vortex arrays~\cite{1979Yarmchuk}. For more details on the original studies related to the production of quantized 
vortices in superfluid helium, we have the 1991 Donnelly text book~\cite{1991Donnelly}, as well as some 
review texts on superfluidity and quantum hydrodynamics, such as Refs.~\cite{2013Tsubota,2013Tsubota-book}.
The more recent theoretical and experimental investigations related to the dynamical formation of vortices in BEC 
system,  as well as regarding to the induced mechanisms to generate them, can be traced by several related reviews, 
as Refs.~\cite{2001Fetter,2005Zwierlein,2008Cooper,2008Bloch,2009Fetter}, which describe the advances
following the first experiments.
 The techniques, such as rotating the magnetic trap, laser stirring, and oscillating excitation superimposing the 
 trapping  potential, have been used to nucleate vortices in BECs~\cite{2000MadisonPRL,2000Chevy,2009Fetter}. 
 In Ref.~\cite{2005Parker}, rotating BEC was studied by considering weak time-dependent elliptical deformation. 
 First, it was shown that due to energy transfer, a turbulent state is produced by the quadrupole instability frequency. 
 Such state can be subsequently damped by vortex-sound interactions leading to a vortex lattice. The classical and 
quantum regimes of two-dimensional turbulence, produced in trapped BEC was studied in 
Refs.~\cite{2012Bradley,2012Reeves,2014Reeves,2014Billam,2019Gauthier},  by exploring spectral kinetic energy analysis associated with the vortex patterns.  From some recent reviews, as Refs.~\cite{2016Tsatsos,2019Madeira,2020Madeira},
we can also trace the present status of quantum turbulence studies associated to quantum gases.
More recently, the fundamental differences between quantum and classical turbulences in rotating systems are 
investigated in Ref.~\cite{2022Estrada}. As considering binary systems, one should notice some recent studies in  Refs.~\cite{2022Dastidar,2022Das} reporting on the dynamics behind vortex-pattern formations. 
Still, more investigations are required in order to understand vortex mechanisms,  and their possible relation with 
sound-wave excitations, which emerge at low-momentum phonon-like behaviors of dilute BEC systems~\cite{2002Pethick,2003Stevenson}.
In a disordered two-dimension (2D) turbulent quantum fluid, it was shown in Ref.~\cite{2014Reeves} the
emergence of coherent vortex structures obeying the Feynman rule~\cite{1955Feynman} of constant average areal 
vortex density.   By considering dipolar BEC systems, the three-dimensional structures of vortices 
were also studied in Ref.~\cite{2016Kishor}, by considering fully anisotropic traps, with increasing eccentricity. 

The production of vortices in multi component BECs are more intriguing due to the diverse vortex lattice 
phases, which can be observed in addition to the more fundamental Abrikosov triangular kind of patterns,
first predicted in type II superconductors~\cite{1957Abrikosov}.  The relevance of vortices generation can 
be traced from the detailed exposition by Abrikosov in his Nobel Lecture~\cite{2004Abrikosov}. His derivation 
of the vortex lattice was done in 1953, being published only when Landau became convinced of the vortices idea 
in 1957,  after the Feynman's contribution on vortices in superfluid helium~\cite{1955Feynman}.
For a more updated overview  of vortex lattice theory focusing related experiments in Bose-Einstein condensates,  
see Ref.~\cite{2009Newton}. 

For the possibilities to study the dynamics of vortex productions in binary mixtures, 
one should also consider the actual control in the production of different BEC atomic species.
With two different hyperfine states of the same rubidium isotope, $^{87}$Rb, 
the production of a BEC binary mixture was first reported in Ref.~\cite{1997Myatt}, 
followed by studies on BEC collisions with collective oscillations in Ref.~\cite{2000Maddaloni}, 
as well as other investigations concerned atomic and molecular 
properties of coupled BEC systems. Among them, we have the first studies with dipolar-dipolar 
interactions in quantum gases being reviewed in Ref.~\cite{2009Lahaye}.
With two different isotopes of rubidium, $^{85}$Rb and $^{87}$Rb, some prospective
theoretical studies done in Ref.~\cite{1998Burke} have already indicating that a stable 
mixed-isotope double condensate could be formed by sympathetic evaporative cooling, 
allowing partial control of the interactions between hyperfine states.
As related to condensed two atomic species,  it was reported in Ref.~\cite{2002Modugno} the 
production of potassium-rubidium ($^{41}$K and $^{87}$Rb) coupled mixture, having
their collisional properties further studied in Ref.\cite{2002Ferrari}, with the control on 
the interspecies interactions explored in Ref.~\cite{2008Thalhammer}.
The experimental observation of controllable phase separation of coupled systems with two species 
of rubidium isotopes, $^{85,87}$Rb, was also reported in Ref.~\cite{2008Papp}.
Following that, it was 
reported several other studies with the production of dual BEC mixtures, in which the  $^{87}$Rb is 
coupled to another atomic species, as with caesium $^{133}$Cs~\cite{2011McCarron}, 
with strontium isotopes $^{84,88}$Sr~\cite{2013pasquiou}, with potassium $^{39}$K~\cite{2015Wacker}
and with sodium $^{23}$Na~\cite{2016Wang}.
More recently, experiments with strongly dipolar ultra-cold gases,  formed with dysprosium (Dy) 
and erbium (Er), have been reported in Refs.~\cite{2018Trautmann,2019Natale}, from where other 
recent references concerned dipolar mixtures can also be traced. 

In view of the actual experimental control of two-component coupled binary systems, it is natural
to look for studies with rotating BEC mixtures. We notice that such studies have mainly been 
concentrated on ground-state vortex structures regarding the miscibility, which is a relevant 
characteristics of multi-component ultra-cold gases.
Their miscibility behavior depends on the nature of the interatomic interactions between different 
species. Miscible or immiscible two-component BEC systems can be distinguished by the spatial 
overlap or separation of the respective wave-functions of each component.
By assuming a binary dipolar mixture, rotational properties were studied in concentrically coupled 
annular traps in Ref.~\cite{2015Zhang}. 
Following previous studies with vortices in dipolar BECs~\cite{2009Wilson,2012Wilson,2016Zhang},
the miscible-immiscible transition of dipolar mixtures with dysprosium ($^{162,164}$Dy) 
and erbium ($^{168}$Er) isotopes,  were studied in Ref.~\cite{2017Kumar}, within pancake- and 
cigar-type geometries, by playing with the trap parameters, together with the two-body scattering 
lengths.  As considering rotating BEC binary mixtures, vortex patterns were explored by some of us in 
Refs.~\cite{2017KumarMalomed,2018Kumar}.  In Ref.~\cite{2019Kumar}, it was also
revealed the mass-imbalance sensibility of a dipolar mixtures, and also shown that the dipole-dipole 
interactions, going from attractive to repulsive, can be instrumental 
for the spatial separation of rotating binary mixtures. In an extension of this study, it was also 
pointed out in Ref.~\cite{2020Tomio} the effect on the rotational properties of a quartic term, 
added to the harmonic trap of one of the components.

Our aim in the present work is to analyze the route to the production of vortices related to 
stirring elliptical trap mechanisms in mass-imbalanced binary coupled BECs, by following the 
associated energetic effects due to low-momentum sound-wave excitations.
The systems will be exemplified by the two experimentally accessible coupled mixtures, given
by rubidium-caesium $^{85}$Rb-$^{133}$Cs,  together with $^{85}$Rb-$^{87}$Rb. These
choices for the binary systems (with large and small mass differences) were motivated by 
verifying possible mass-imbalanced effects.
To understand the mechanism of vortex generation, we analyze the corresponding kinetic 
energy parts of both components of the mixture, which are expected to be clearly show up
for large mass differences. Following this approach, from the corresponding averaged classical 
angular momentum and torque obtained for each of the components, we estimate a rotational 
velocity in order to produce simulations in the rotating trap (when the stirring potential is
removed).  By assuming this classical effective rotational frequency for the rotating magnetic trap, 
the ground-state of the system is obtained, together the corresponding vortex patterns for 
the densities. 

In the following Sect.~\ref{sec2}, split in four parts, we present the necessary details on the 
mean-field  formalism for rotating binary BEC system, in which the trap interaction is composed by
the usual 2D pancake-like trap, which can be rotated, together with a time-dependent 
stirring potential.  The time-dependent energetic relations for the stirring model are provided in 
Sect.~\ref{sec2}B. The corresponding classical time-averaged rotational velocities for a rotating 
model without the stirring potential are also given in this section, followed by the structure factor 
expression associated to the lattice order. 
In Sect.~\ref{sec2}C, we include the formalism concerned the results obtained
for the kinetic energy spectra in rotating-frame field. Numerical details and parameters are 
presented in Sect.~\ref{sec2}D.
For the main results, we have two sections: The Sect.~\ref{sec3} provides details on the 
time evolution of observables related to the two-kind of mass-imbalanced coupled systems 
we are studying, considering the stirring interaction. The final vortex pattern 
results are also compared with ground-state results, obtained when the stirring potential 
is replaced by the corresponding classical rotational frequency. 
In Sect.~\ref{sec4}, we perform an analysis related to the kinetic energy spectra.
The main focus in this case, was to verify the behavior of the incompressible kinetic energy spectra 
in different intervals of the time evolution, before the vortices nucleation till stable vortex patterns
are established. In both Sects.~\ref{sec3} and \ref{sec4}, relevant aspects due to mass differences 
in the mixtures are discussed concerned the vortex patterns and energetic distribution.
Our general conclusions and perspectives are summarized in Sect.~\ref{sec5}.

\section{Stirred BEC coupled formalism}\label{sec2}
In our present approach, we assume a binary coupled system, in which the two atomic species 
$i=1,2$ have non-identical masses $m_i$. 
We further assume that both species have identical number of atoms $N_i\equiv N$
and are confined by strongly pancake-shaped harmonic traps, with longitudinal and transversal 
frequencies given, respectively, by $\omega_{i,z}$ and $\omega_{i,\perp}$, with
fixed aspect ratios given by $\lambda\equiv \lambda_i=\omega_{i,z}/\omega_{i,\perp}=10$.
As our study will be mainly concerned with the analysis of dynamical occurrence of vortices in 
an originally stable miscible binary system, the intra-species scattering lengths are assumed
to be fixed and identical, $a_{11}=a_{22}=60a_0$ (where $a_0$ is 
the Bohr radius), such that the relative strength is controlled by the inter-species scattering 
length $a_{12}$. Also to guarantee that the coupled system is in a miscible state, 
we assume $a_{12}/a_{ii}=1/2$, the inter-species interaction being about half of the 
intra-species ones. These assumptions related to the non-linear interactions relies on the 
actual experimental possibilities to control the interactions via Feshbach resonance 
mechanisms~\cite{1999Timmermans}.

The coupled Gross-Pitaevskii (GP) equation is cast in a dimensionless format, with the energy and 
length units given, respectively, by $\hbar \omega_{\rho}$ and 
$l_{\rho}\equiv \sqrt{\hbar/(m_1\omega_{\rho})}$, in which we are assuming the first species (the 
less-massive one) as our reference in our unit system, such that $\omega_\rho\equiv\omega_{1,\perp}$
Correspondingly, the full-dimensional space $\bf r$ and time $t$ variables are, respectively, replaced by 
dimensionless ones. For convenience, we kept the same representation, ${\bf r}/l_\rho\to {\bf r}$ and 
$\omega_{\rho}t\to t$, with the new variables understood as dimensionless. 
 Within these units,  for simplicity, we first adjust the confining trap frequencies of both species with 
${m_2\omega_{2,\perp}^2}={m_1\omega_{\rho}^2}$,
 such that the dimensionless non-perturbed three-dimensional (3D) traps for both species can be represented by the 
 same expression, given by 
\begin{equation}
V_{i,3D}({\bf r})=\frac{1}{2}(x^{2}+y^{2}+\lambda^2 z^2) \equiv V_0(x,y) + \frac{1}{2}\lambda^2 z^2,
\label{3Dtrap}
\end{equation}
where $V_0(x,y)$ is the 2D harmonic trap. In this way, there is no explicit 
mass-dependent factor in the trap potential, which will remain only in the kinetic energy term.  
By fixing to very large values the aspect ratio $\lambda$, we can follow the usual approximation, 
in which we have the factorization of the total 3D wave function, 
$\psi_{i}(x,y,t)\chi_i(z)$, with 
$\chi_i(z)\equiv \left(\lambda_i/{\pi}\right)^{1/4}e^{-{\lambda_i z^{2}}/{2}}$. 
In this case, the ground-state energy for the harmonic trap in the $z-$direction becomes a constant 
factor to be added in the total energy. 
It is safe to assume a common mass-independent transversal wave-function for both components, with 
$\lambda_i=\lambda$, as any possible mass dependence can be absorbed by changing the corresponding aspect ratio.
This approach for the reduction to 2D implies that we also need to alter the nonlinear parameters accordingly, as the integration on the $z-$direction will bring us a $\lambda-$dependence in the 
non-linear parameters.  

\subsection{Gross-Pitaevskii coupled stirred model}
Motivated by some previous 
studies in which stirring potentials have been used to single condensed atomic 
species~\cite{2000MadisonPRL,2001Hodby,2001Shaeer},
in our study we consider that the binary coupled system can be further perturbed by the action of a 
laboratory-frame time-dependent stirring potential~\cite{2005Parker}, given by
{\small\begin{equation}
V_s(x,y,t)= \frac{\epsilon}{2} \left[(x^2-y^2) \cos(2\Omega{_E} t)
-2xy\sin(2\Omega{_E} t)\right],
\label{Vs}
\end{equation}
}where $\Omega_E$ (given in units of $\omega_\rho$) is an elliptical stirring laser frequency, with 
$\epsilon$ the corresponding strength. 
Our approach to generate patterns of vortices in both coupled condensates is initially fully based on the 
stirring potential.
However, for a further correspondence analysis, we also consider the possibility to produce the same 
pattern of vortices in the ground state, with the replacement of the stirring time-dependent potential 
by a constant frequency $\Omega_0$ (units $\omega_{\rho}$), for each component of the coupled system.
Therefore, from the above, with the trap potentials given by Eqs.~\eqref{3Dtrap} and \eqref{Vs}, the 
corresponding coupled 2D GP equation for the two-components wave functions, $\psi_{i}\equiv \psi_{i}(x,y,t)$, 
normalized to one, $\int_{-\infty}^{\infty}dx\int_{-\infty}^{\infty}dy\,|\psi_{i}|^{2}\equiv
\int d^2r\,|\psi _{i}|^{2}=1,$ is given by
{\small\begin{eqnarray}
\mathrm{i}\frac{\partial \psi_{i}  }{\partial t }&=&\left[\frac{-m_{1}}{2m_{i}}{\nabla^2
}+V_0(x,y)+\sum_{j}g_{ij}|\psi_{j} |^{2}\right]
\psi_{i}  \nonumber \\
&+&\left[V_s(x,y,t)- \Omega_0 L_{z}\right]\psi_{i}
\label{2d-2c}, 
\end{eqnarray}
}where $-{\rm i}\nabla \equiv -{\rm i}\left(\hat{e}_x\frac{\partial}{\partial x}+\hat{e}_y\frac{\partial}{\partial y}\right)$
is the 2D momentum operator, with $g_{ij}$ being 
the two-body contact interactions, related to the intra-species and inter-species 
scattering lengths, respectively, $a_{ii}$ and $a_{ij} (j\ne i)$. It is given by
\begin{eqnarray}
g_{ij}\equiv \sqrt{2\pi\lambda}
\frac{m_1 a_{ij} N_j}{\mu_{ij}l_\rho},
\label{par}
\end{eqnarray}
where $\mu_{ij} \equiv m_im_j/(m_i+m_j)$ is the reduced mass. 
In the next, for our numerical simulations, the length unit will be adjusted to  
$l_\rho = 1\mu$m$ \approx 1.89\times 10^4 a_0$, such that $a_{ij}$ can be conveniently 
given in terms of $a_0$.
For the additional trap potential, which appear  in Eq.~\eqref{2d-2c}, we have the stirring 
time-dependent potential $V_s(x,y,t)$, given by Eq.~\eqref{Vs},  and $\Omega_0 L_z
\equiv -{\rm i}\Omega_0 \left(x\frac{\partial}{\partial{y}}-y\frac{\partial}{\partial{x}}\right)
={\rm -i}\Omega_0\frac{\partial}{\partial\theta}$.
Here, and whenever convenient, we also consider the usual polar coordinates, 
with $x=r\cos\theta$, $y=r\sin\theta$ ($0\le r<\infty$, $0\le\theta\le2\pi$) and 
$r^2\equiv x^2+y^2$. 

From Eqs.~\eqref{3Dtrap} and \eqref{Vs}, the total 2D trap potential, with the time-dependent
stirring term, having identical form $V_i(x,y,t)\equiv V_0(x,y)+V_s(x,y,t)$ for both components 
$i$, can be expressed in cylindrical coordinates, as
\begin{eqnarray}
V(r,\theta,t)&=&
\frac{r^2}{2}\left[1+\epsilon\;\cos\left(2\theta+2\Omega_E t\right)\right],
\label{2Dtrap}
\end{eqnarray}
where a small value $\epsilon=0.025$ is used to maintain the condensate with an
approximate symmetric 2D shape during the time evolution~\cite{2005Parker}.
As the laser stirring is more helpful to rotate BECs for values greater than the trap frequency,  
in our simulations, in units of the trap frequency $\omega_\rho$, we assume $\Omega_E$ =1.25.
 In the long-time evolution of the coupled system, this stirring frequency should provide 
vortex patterns in each of the components of the mixture.

In our following approach, we first explore the vortex-production case provided by the time-dependent
stirring potential, with $\epsilon\ne 0$ and $\Omega_0=0$, such that the only rotation will be derived from the 
stirring potential, induced by the frequency $\Omega_E$.
As it will be shown, this will result in contributions of sound waves as well as vortices, 
which will be further explored by the analysis of the associated observables, within quantum
considerations of the kinetic energies.
 The results obtained by the stirring potential are followed by a classical analysis 
considering the rotational velocity, which can provide an effective frequency $\Omega_{cl}$. 
Therefore, with $\Omega_0=\Omega_{cl}\ne 0$ in Eq.~\eqref{2d-2c}, a correspondence is established 
between the stirring procedure to generate the vortex patterns with another direct approach in which 
the patterns are verified without the stirring time-dependent potential.

\subsection{Dynamics of stirred vortex formation}
In general, quantum gases are compressible fluids, such that their corresponding density can change
when submitted to a force. This is true to a certain degree, as part of the fluid can behave as an 
incompressible fluid, similar as a liquid.
In our present case, the condensate is submitted to a time-dependent stirring potential,
associated to a torque, which is mainly due to a part of the rotational kinetic energy, that
we can call as the compressible one.
Therefore, for the analysis of this behavior,  we start by considering the total energy, in which
only the harmonic trap with the time-dependent stirring potential are assumed for the total
trap potential, such that $\Omega_0=0$ in Eq.~\eqref{2d-2c}.  So,  for each component of the 
mixture, the total energies $E_i(t)$ are given by
{\small\begin{eqnarray}
E_i(t)&=&
\int d^2{\bf r} \left[\frac{m_{1}}{2m_{i}}|\nabla\psi_i |^2 + V_i(x,y,t)n_i(x,y,t)\right] 
\nonumber\\&+&\frac{1}{2}\sum_{j=1,2}g_{ij} \int d^2{\bf r}\; n_i(x,y,t) n_j(x,y,t)
\label{En}, 
\end{eqnarray}
}where $n_{i=1,2}(x,y,t)\equiv |\psi_i|^2$ are the respective time-dependent densities. 
For each species, with
the current densities ${\bf j}_i(x,y,t)$ expressed in terms of the respective densities and
velocity fields ${\bf v}_i(x,y,t)$, such that ${\bf j}_i(x,y,t)=n_i\; {\bf v}_i(x,y,t)$,  we can write
\begin{equation}\label{current}
{\bf v}_i(x,y,t) = \frac{1}{2\mathrm{i}\; |\psi_i|^2} { \big[ \psi_i^\star \nabla\psi_i - 
\psi_i \nabla \psi_i^\star \big]}.
\end{equation}
The associated kinetic energies,
\begin{align}\label{Ken}
E^K_i(t) = \frac{m_1}{2m_i} \int d^2{\bf r}\,n_i(x,y,t) |\mathrm {\bf v}_i(x,y,t)|^2,
\end{align}
can be decomposed  in compressible $E_i^{c}(t)$ and incompressible $E_i^{nc}(t)$ parts. 
For that, the component $i$ of the density-weighted velocity field, defined as 
${\bf u}_i(x,y,t)\, \equiv\,\sqrt{n_i(x,y,t)} {\bf v}_i(x,y,t)$, can be split into 
an incompressible part, 
${\bf u}_i^{(nc)}\equiv {\bf u}_i^{(nc)}(x,y,t)$ satisfying $\nabla . {\bf u}_i^{(nc)} = 0$, 
and a compressible one, ${\bf u}_i^{(c)}\equiv {\bf u}_i^{(c)}(x,y,t)$, satisfying 
$\nabla \times{\bf u}_i^{(c)} = 0$. 
Therefore, with
${\bf u}_i(x,y,t)\, =\, {\bf u}_i^{(nc)}(x,y,t) + {\bf u}_i^{(c)}(x,y,t)$, the kinetic energies
are split as 
{\small\begin{align}\label{Ken-split}
E_i^K(t)&=E_i^{nc}(t)+E_i^{c}(t)\nonumber \\
&\equiv \frac{m_1}{2m_i} \int d^2{\bf r} \left[|{\bf u}_i^{(nc)}|^2 +
|{\bf u}_i^{(c)}|^2 
\right].\end{align}
}In 2D momentum space, with ${\bf k}=({k_x},{k_y})$ and $d^2k=dk_xdk_y$, these two
parts of the kinetic energies, with $(\alpha)\equiv (c), (nc)$, can be written as
\begin{eqnarray}\label{Ken-alpha}
E_i^{(\alpha)}(t) &=& \frac{m_1}{2m_i} \int d^2{\bf k}|\mathcal{F}_i^{(\alpha)}({\bf k},t)|^2,\\
&=& \frac{m_1}{8\pi^2m_i} \int d^2{\bf k}\left|   \int d^2{\bf r}\, e^{-{\rm i}{{\bf k}.{\bf r}}} {\bf u}_i^{(\alpha)}
\right|^2
\nonumber.\end{eqnarray}
To analyze the contributions of sound waves as well as vortices, both compressible 
and incompressible parts of the kinetic energies are determined.
The torque experienced by the time-dependent stirring potential,  $V_s(r,\theta,t)$,  
which is given by  Eq.~\eqref{2Dtrap}, can be obtained through the corresponding torque 
operator,  
\begin{eqnarray}\label{torque}
{\bf \tau}_z ({\bf r},t)&=& 
-\frac{\partial}{\partial\theta}
V_s(r,\theta,t) = \epsilon\; {r^2}\sin(2\theta+2\Omega_E t),
\end{eqnarray}
which corresponds to apply a rotation in the elliptical time-dependent part of the potential, 
with $2\Omega_E t \to 2\Omega_E t - \pi/2$.
Due to this change in the time-dependent potential, it follows that the expected values of the 
induced angular momenta,  $\langle L_z(t)\rangle_i$, and respective moment of inertia, 
$\langle I(t)\rangle_i$,  are   given by 
\begin{eqnarray}
\langle L_z(t)\rangle_i &=& {-\rm i}
\int d^2r \;\psi_i^\star 
\frac{\partial}{\partial\theta}\psi_i \;\;{\rm and}\;\;
\langle I(t)\rangle_i =\int d^2r |\psi_i|^2\;r^2,\nonumber \end{eqnarray}
with the associated classical rotational velocity being
\begin{equation}
\Omega_i(t)\equiv \frac{\langle L_z(t)\rangle_i }{\langle I(t)\rangle_i }.
\label{rot-cl}\end{equation}  
We are aware that it is a non-trivial problem to calculate the moment of inertia of a superfluid or a
condensate system. 
In the literature, the perturbation theory is commonly used to calculate the momentum of inertia for 
trapped ideal Bose gases. It is a temperature dependent treatment, in which the condensed and 
non-condensed parts are considered into the calculation of the moment of inertia of a quantum 
system. Since we are working with a pure GP model, the non-condensate thermal clouds are not 
taken into account. We just use the classical relation to obtain approximate rotational velocities.. 
 
 The structure factor provides us
information about the periodicity of the condensate density. For triangular
lattice geometry, there are periodic peaks of regular hexagonal geometry. By
looking at the position of peaks of the structure factor, we can distinguish
between the Abrikosov lattice and the pinned vortex lattice

We can also consider for the coupled system the density structure factor~\cite{2007Sato}, which is a 
function of the $k\equiv \sqrt{k_x^2+k_y^2}$ that provides information on the lattice order, the periodicity of
the condensate density. This structure factor can be expressed by
\begin{eqnarray}\label{sf}
s_i(k) &=& \frac{1}{2\pi}\int d^2r\; |\psi_i|^2 e^{-{\rm i}{\bf k.r}}.
\end{eqnarray}
 
\subsection{Kinetic energy spectrum in rotating-frame field}
The rotating-frame velocity field, at a given frequency $\Omega$, is defined by subtracting the
corresponding rigid-body velocity field, as 
\begin{align}\label{vrot}
    \mathbf{v}_{\Omega}&=\mathbf{v}-(\mathbf{\Omega}\times\mathbf{r})_z,
\end{align} 
such that, for each density-component $n_i$, the associated particle current
$\mathbf{j}_{i,\Omega}({\bf r},t)\equiv n_i({\bf r},t)\mathbf{v}_{i\Omega}({\bf r},t)$ satisfies the continuity equation
\begin{align}
    \frac{\partial n_i({\bf r},t)}{\partial t}+\nabla\cdot\mathbf{j}_{i\Omega}({\bf r},t)=0.
\end{align} 
Therefore, by following Eqs.~\eqref{Ken} to \eqref{Ken-alpha}, when considering a rotating frame, 
at a given frequency $\Omega$, we obtain
\begin{align}\label{spec1}
E^{(\alpha)}_{i,\Omega}(t)&= \frac{m_1}{2m_i} \int d^2{\bf k} |\mathcal{F}^{(\alpha)}_{i,\Omega}
({\bf k},t)|^2\nonumber\\
&\equiv \frac{m_1}{m_i}\int_0^\infty dk\;E^{(\alpha)}_{i,\Omega}(k,t),
\end{align}
which is defining the  
\emph{velocity power spectral density} in $k$ space as given by 
\begin{align}\label{spec}
E^{(\alpha)}_{i,\Omega}(k,t) &= {k} \int_0^{2\pi}d\theta_k |\mathcal{F}^{(\alpha)}_{i,\Omega}({\bf k},t)|^2.
\end{align}
Within this definition, for convenience, we are removing the relative mass ratio difference, which does not 
affect the corresponding spectral behavior of each component.
This equation is used to calculate the kinetic-energy (or velocity) power spectra densities presented in this work. 
 The computational methods for calculating velocity power spectrum using decomposed kinetic energies 
are discussed in more detail in Ref.~\cite{2021Bradley}.
 
\subsection{Numerical approach and parameters} 
In our approach to obtain the numerical results presented in the next section, we solve the 2D coupled 
GP mean-field equation by the split time-step Crank-Nicolson method, with an appropriate algorithm for
discretization as prescribed in Ref.~\cite{2019KumarCPC}. Within our dimensionless units, in which the
time unit is the inverse of the transversal trap frequency $1/\omega_\rho$ and we
fix the length unit at $l_\rho = \sqrt{\hbar/(m_1\omega_\rho)}=1\mu$m$ \approx 1.89\times 10^4 a_0$,
we consider the space step sizes as 0.05, with time steps $10^{-4}$. 
The chosen space and time steps are verified to be enough smaller to provide stable results along the
time evolution of the coupled system. The space steps are more than four times smaller than 
the healing lengths, which are the cut-off lengths at the vortex core sizes.

With the binary mixture considered in a completely miscible state, with identical intra-species scattering
lenghts ($a_{11}=a_{22}$), the miscibility parameter $\delta\equiv a_{12}/a_{11}$ must be smaller than 1.
For that, we assume that both intra-species scattering lengths are repulsive, with $a_{ii}=60a_0$, being 
twice the inter-species one $a_{12}=30a_0$, such that $\delta=0.5$.  
We also consider both species with identical number of atoms, $N_{1}=N_{2}=10^4$.  
With these parameters, we observe that the miscible-state condition $\delta=0.5$ will correspond to 
$g_{12}/g_{11}$, given by~\eqref{par}, slightly smaller due to the mass differences of the considered
binary systems.

The healing lengths are associated to the dimensionless chemical potentials $\mu_i$ 
(units $\hbar\omega_\rho$), which are obtained from the stationary solutions, 
$\psi_i=\sqrt{n_i}e^{-{\rm i}\mu_i t}$, given by Eq.~\eqref{2d-2c}, in which $\Omega_0$ and the stirring 
potential are set to zero.  So, with energies in units of $\hbar\omega_\rho$, the healing lengths are in units of 
$l_\rho=1\mu$m, with the speed of sound in units of $l_\rho \omega_\rho$. 
For the  binary mixture  $^{85}$Rb-$^{133}$Cs, the respective chemical potentials are 
$\mu_1=\mu_{^{85}\text{Rb}}=15.10$ and 
$\mu_2=\mu_{^{133}\text{Cs}}=12.87$. Correspondingly, the healing lengths $\xi_i$ and 
sound velocities $v_i$ are
$\xi_1={\xi}_{^{85}\text{Rb}} =0.257$, 
$\xi_2={\xi}_{^{133}\text{Cs}} =0.223$
$v_{s,1}={v}_{^{85}\text{Rb}}=3.43$ and 
$v_{s,2}={v}_{^{133}\text{Cs}}=3.71$.
In the case of the almost identical-mass mixture, $^{85}$Rb-$^{87}$Rb, these values are given by
$\mu_1=\mu_{^{85}\text{Rb}}=15.53$,
$\mu_2=\mu_{^{87}\text{Rb}}=15.40$,
$\xi_1={\xi}_{^{85}\text{Rb}}=0.254$, 
$\xi_2={\xi}_{^{87}\text{Rb}} = 0.252$,  
$v_1={v}_{^{85}\text{Rb}}\approx v_2={v}_{^{87}\text{Rb}}=4.04$.
Noticed that, due to our choice of units based on the first component,
 the corresponding full-dimensional expressions of the healing lengths are given by 
 $\tilde{\xi}_i= l_\rho\sqrt{m_1/(m_i\mu_i)}\equiv l_\rho \xi_i$, with 
 the second component carrying a mass factor in its relation with the chemical potential.

In the next section, we provide details on our main results for the time evolution of the coupled
condensates. They are the vortex-pattern densities, formed after long-time evolution of the system,
together with other associated observables, as the compressible and incompressible kinetic energies, 
the current densities and torques.  From the time-dependent stirring potential, averaged 
effective classical rotational frequencies and structure factors are obtained, which are further considered 
in a comparative analysis with ground-state rotating solutions, without the stirring potential.

\section{Vortex-pattern distribution results}\label{sec3}
In our study we have considered two mass-imbalanced mixtures, such that we can contrast the results
obtained by a mixture with large mass difference between the components, given by  $^{85}$Rb$-^{133}$Cs,
with the results obtained by a system with negligible mass difference, exemplified by $^{85}$Rb$-^{87}$Rb.
Once, for both the cases, stable lattice patterns are verified, averaged frequencies are derived from the 
results given by the stirring model, which are further considered to obtain the lattice patterns from the 
ground-state time-independent results.
 
 Therefore, we split our results in two parts, corresponding to the two binary systems we are studying.

\subsection{Binary $^{85}$Rb$-^{133}$Cs mixture}
In this subsection we are presenting the main results considering the binary mass-imbalanced system 
composed by the $^{85}$Rb and $^{133}$Cs. As previously stated, the sample results we have obtained
are for fixed scattering lengths in a miscible configuration, such that 
$\delta = 0.5$.  In view of the actual experimental control on the condensation of these two atomic
species, we understand that Feshbach resonance techniques~\cite{1999Timmermans} can be considered 
to approximately bring the coupled system to this hypothetical condition. 

\begin{figure*}[htbp]
\begin{center}
\includegraphics[width=0.95\textwidth]{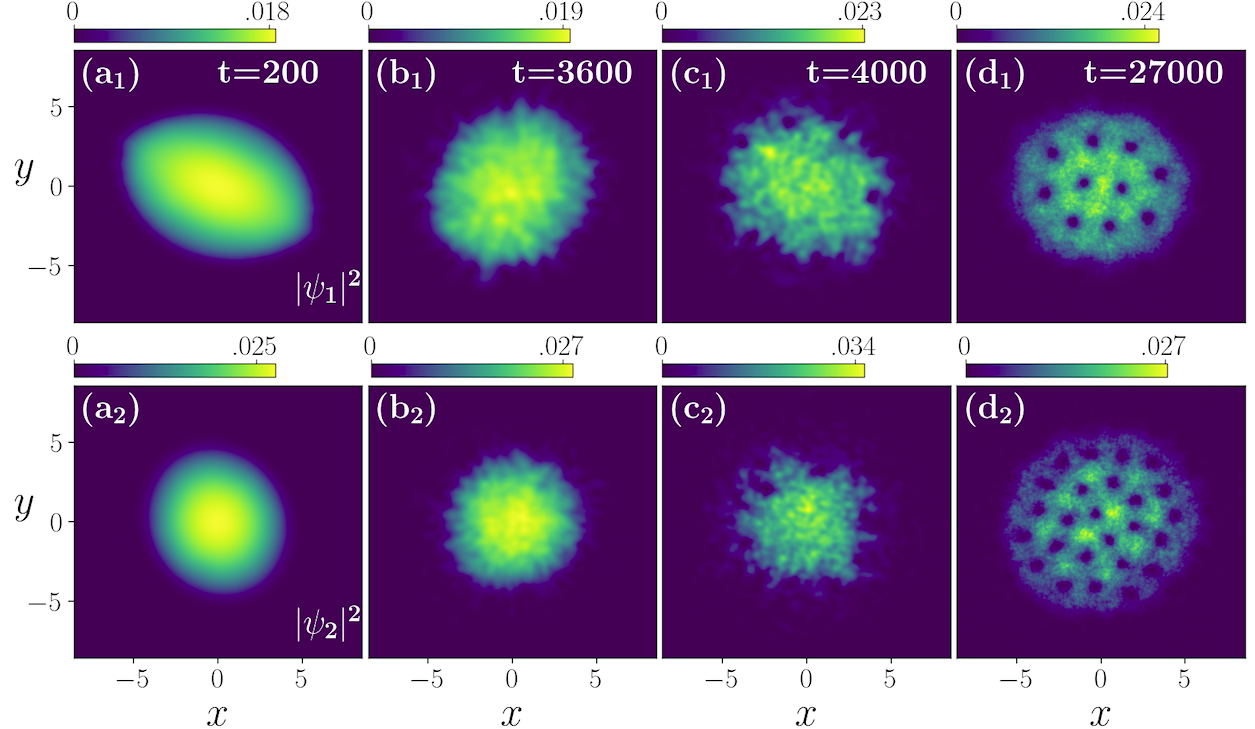}
\end{center}
\caption{(Color on line) This set of panels is shown the time evolution of the densities for both condensed species 
of mixture $^{85}$Rb-$^{133}$Cs. The upper panels are for the $^{85}$Rb (component 1), with the lower panels 
for $^{133}$Cs (component 2). The system is in a miscible configuration with $\delta=0.5$ (scattering lengths are 
$a_{ii}=60 a_0$ and $a_{12}=30 a_0$).  Starting with a ground-state solution, the system is evolved in real time with 
$\Omega_E=1.25$ with $\epsilon=0.025$.  The density levels are indicated at the top of each panel.
All quantities are dimensionless, with $l_\rho$ and $\omega_\rho$ being the respective length and frequency units. 
}
\label{f01}
\end{figure*} 
In Fig.~\ref{f01}, through the time evolution of the densities, we have the main results for the densities, in which
it was shown the route to the vortex lattice generated by the stirring potential. 
By considering four set of panels, within the 2D space defined by the cartesian coordinates (units of $l_\rho$),
we provide the results for the densities at some specific time intervals, characterizing specific regimes,
for both $^{85}$Rb (upper panels, with labels 1) and $^{133}$Cs (lower panels, with labels 2) species. 
In order to obtain stable non-rotating ($\Omega_0=0$) ground-state solutions, we first solve  the coupled 
Eq.~\eqref{2d-2c} by using imaginary time ($t\rightarrow -\mathrm{i}t$), with  $\epsilon=0$ (without the 
stirring potential).  Next, this ground-state solution is evolved in real time, by considering a rotating 
elliptical frequency $\Omega_E=1.25$ with the strength of the stirring potential $\epsilon=0.025$, as given 
in Eq.~\eqref{2Dtrap}. The first coupled panels of Fig.~\ref{f01}, (a$_1$) and (a$_2$),  are for $t=200$,
which refers to the time interval when shape deformations start to occur in the system. As verified, the 
condensed system is already displaying elliptical deformations (quadrupole excitations), due to 
the stirring potential, which periodically reverse back to the symmetric 2D trap, with the expected full cycle given by 
$\pi/\Omega_E$.  The shape deformation is more noticeable in the first lighter component ($^{85}$Rb) than in the 
second one ($^{133}$Cs).  This periodic behavior goes until a break occurs in the symmetry, with vortex 
nucleations happening at the surface of the condensed species, after enough long-time evolution 
($t > 3000$). For this second interval, just before and at the time the nucleation of vortices start to appear, we select
two sets of plots in Fig.~\ref{f01}, for $t=3600$ [panels b$_1$) and (b$_2$)] and for $t=4000$  [panels c$_1$) and (c$_2$)].

The time interval, in which we can observe the break in the symmetry, is being 
represented by the two set of panels (b$_i$) and (c$_i$), within an interval of $\Delta t = 400$ 
(with $t=$3600 and 4000).  As the time goes increasing, after an interval in which turbulent behaviors are 
verified, the nucleated vortices move from the surface to inner part of the condensates, with lattice patterns 
start appearing till the equilibrium is achieved for both coupled condensates.  At the final configuration, we 
noticed that the number of nucleated vortices emerging in the heavier component is significantly larger than  
the number of vortices obtained for the lighter species. 

\begin{figure}[htbp]
\begin{center}
\includegraphics[width=0.5\textwidth]{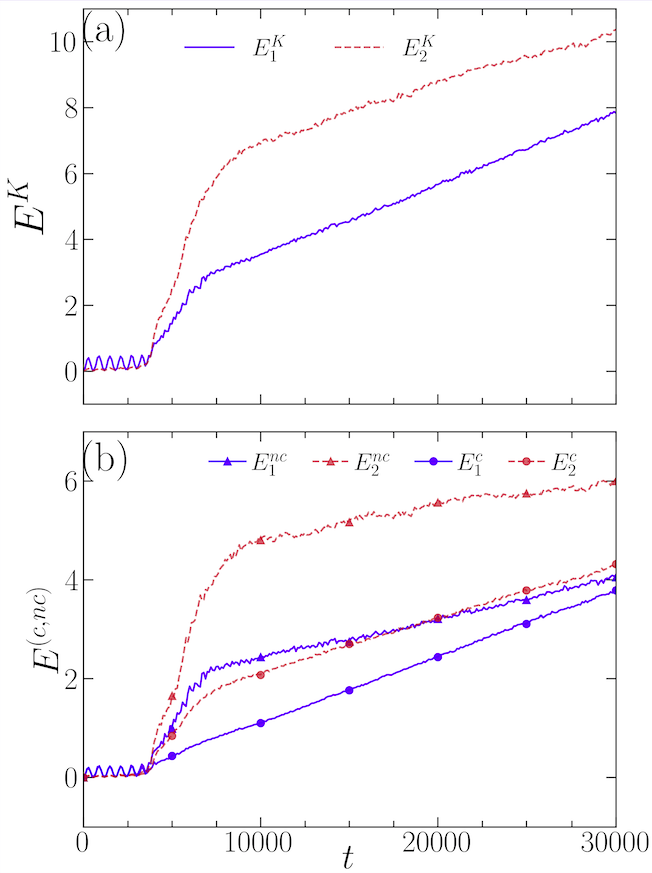}
\end{center}\vspace{-.5cm}
\caption{(Color on line) 
For each component of the mixture, $^{85}$Rb ($i=$1, solid-blue lines) and $^{133}$Cs 
($i=$2, dashed-red lines), the time-evolutions are presented (in the laboratory frame) for 
the kinetic energies, $E^K_i\equiv E^K_i(t)$ [panel (a)] and their corresponding compressed 
($E^{c}_i$, indicated with bullets) and decompressed ($E^{nc}_i$, indicated with triangles) 
parts [panel (b)]. 
With initial configuration being ellipsoidal, the evolution is shown till stable vortex 
patterns are verified.  All quantities are dimensionless, 
with $1/\omega_\rho$ and $\hbar\omega_\rho$ being, respectively, the time and energy units.
}\vspace{-.5cm}
\label{f02}
\end{figure}
 The density evolutions represented in  Fig.~\ref{f01} can be better analyzed by considering the corresponding
 evolution of the associated kinetic energies, which are shown in Fig.~\ref{f02} in the laboratory frame.
 For that, considering the two condensed species, the total kinetic energies are shown in panel (a), which are 
 decomposed in two parts shown in panel (b), one which is compressible and another, incompressible.
The energy oscillations in the initial time interval correspond to the breathing mode oscillations of the condensates. 
The energies display increasing behaviors, from the elliptical deformation phases of the condensates, returning to the 
established energies as they follow the trap symmetry.  The shape deformation is more noticeable in the first component 
($^{85}$Rb) than in the second one ($^{133}$Cs), which can be understood by the different inertia of the species.
The increasing of both component energies occurs due to the vortex nucleations. In this case, we observed that
a more drastic increasing is verified for the second heavier one.
At the final configuration, it is also noticed that the number of nucleated vortices emerging in the condensed 
heavier component is significantly larger than inside the lighter one. 
In general, the energies in rotating frame decrease, when the rotation frequency, or vortex number increases. 
However, in the laboratory frame, the total energy increases regarding vortex number.  The laboratory frame
continuous growing of the energies, verified even after saturation of the vortex number, is associated to the 
applied stirring time-dependent potential.  

To analyze the different dynamical behavior as the energy increases, the kinetic energy is decomposed in two 
parts, considering the compressibility of each condensed parts of the mixture. 
The kinetic energy behaviors of both components are presented in the panel (b) of Fig.~\ref{f02}, 
by considering the previously discussed decomposition of the total kinetic energy [panel  (a)], in an
 incompressible and compressible parts.
The increasing of the energies can be visually verified, in the plots shown in 
Fig.~\ref{f02},  that becomes linear in the long-time interval, as for $t>10^4$. 
It is more significant to analyze the kinetic energy of the system to understand the vortex generation, because 
all the information about the velocity fields can be extracted from the associated kinetic energies, such as the 
elements related to sound-wave propagation and vortex generation. 
Therefore, it is appropriate to distinguish in the total kinetic energy, the parts corresponding to acoustic 
wave propagations (compressible energy) from the ones related to vortex generation (incompressible energy).  
The total kinetic energy contribution has a similar oscillating behavior as the total energy in the initial time 
evolution, with huge energy contributions arising due to vortex and sound-waves productions 
along the long-time dynamical rotation. In this process, the main contributions in the kinetic energies will come 
from the incompressible energies, but having significant contributions also from the compressible parts, as verified in 
the panel (b) of Fig.~\ref{f02}.
By also considering the total energy, in comparison with the total kinetic energy, 
we observe that in the first stage of the time evolution, before the vortices nucleation, the kinetic energy is just a small 
fraction of the total energy, with the oscillating behavior stronger in the lighter component. 
However, in the longer-time evolution, the total kinetic energy of the  $^{133}$Cs component grows to be more 
than 30\% of the total energy; whereas, the corresponding total kinetic energy increasing of the 
 $^{85}$Rb element becoming near 20\% of the total energy.
 This is consistent with the observed number of vortices emerging in the case of $^{133}$Cs, when compared with
 the corresponding number for the  $^{85}$Rb, as verified by the results given in Fig.~\ref{f01}.

\begin{figure}[htbp]
\begin{center}
\includegraphics[width=0.5\textwidth]{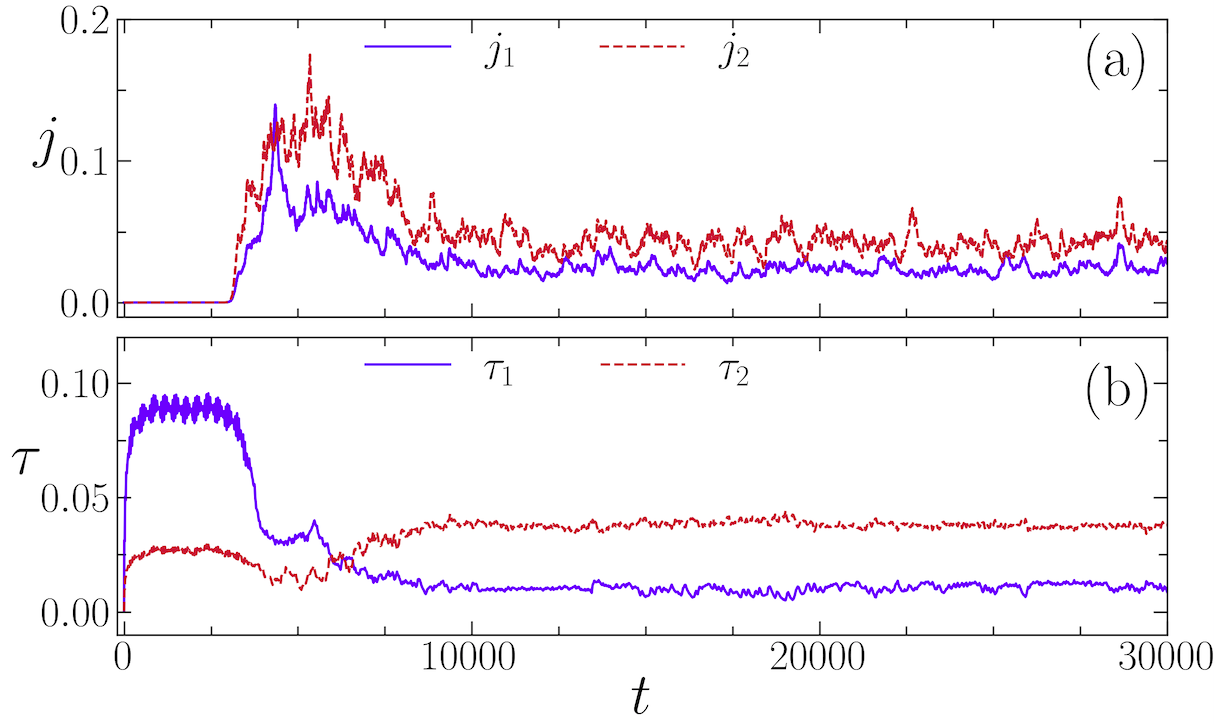}
\end{center}
\caption{(Color on line) For the $^{85}$Rb ($i=1$) and $^{133}$Cs ($i=2$) mixture, 
the respective time evolutions of the current densities [panel (a), given by ~\eqref{current}] and 
torques [panel (b), given by \eqref{torque}] are shown till stable 
vortex patterns are verified. 
All quantities are dimensionless, with ${\omega_\rho^{-1}}$ and $\hbar\omega_\rho$, respectively, being 
the time and energy units.
}
\label{f03}
\end{figure}

The vortex nucleation, responsible for the increasing in the kinetic energies, starts to occur at the surface, 
moving to the inner part of the condensates, with the incompressible parts providing the main contribution 
to the total kinetic energy.  
While comparing both components, the compressible and incompressible parts of the kinetic energy of the 
second component ($^{133}$Cs) are larger than the ones obtained for the first component ($^{85}$Rb),
by a factor that, in a longer-time interval, can be associated to the corresponding mass ratio between the two species:
In order to improve our understanding on the dynamics of the vortex lattice formation, from the rotating
elliptical trap, in this case of large mass-imbalanced 
mixture, we also plot the complete time-evolution of the current density (${\bf j} =n\bf v$) and corresponding
torque for the two species, till the vortex-patterns are generated and become stable. 
The results are presented in Fig.~\ref{f03}, with the current densities shown in the upper frame (a), and the torques 
given in the lower frame (b). 
As verified, in the initial evolution, the current densities are zero for both components of 
the mixture, until a symmetry breaking occurs near $t\sim 3000$, with vortices being nucleated at the surface. 
The current densities reach their maxima around $t\sim 5000$, when the mixture is in a turbulent condition, 
decreasing to an average value consistent with the stabilization of the number of vortices being generated.
Consistently, in all the time evolution, the current is higher for the more massive element of the mixture, the
$^{133}$Cs in the present case. This higher current is consistent with the larger number of vortices being
generated.

In the panel (b) of Fig.~\ref{f03}, in correspondence with the panel (a), the torques experienced by both components, 
due to the stirring potential, are shown during the time evolution of the system.
 The time-dependent stirring trap potential provides the initial rotational energy to the system to rotate. This 
 increasing in the rotational energy is distributed between the two components, being more effective in the case 
 of the lighter component, with the peak in the torque reaching its maximum in the time interval $t<3000$. 
 The heavier element, with corresponding higher inertia, is feeling an initial torque that is about 30\% of the 
 lighter one. However, in the longer time interval, at the equilibrium state, with both averaged values varying 
 close to fixed stable points, it is verified that the heavier element experience higher torque than the lighter one. 
 These results can be associated approximately with the number of vortices induced in the densities of both 
 components shown in Fig.~\ref{f01}. 
 The square-root of the torque ratio is of the same order of the square of the mass ratios, and also the 
 ratio between the number of vortices.
 
\begin{figure}[htbp] 
\begin{center}
\includegraphics[width=8.5cm]{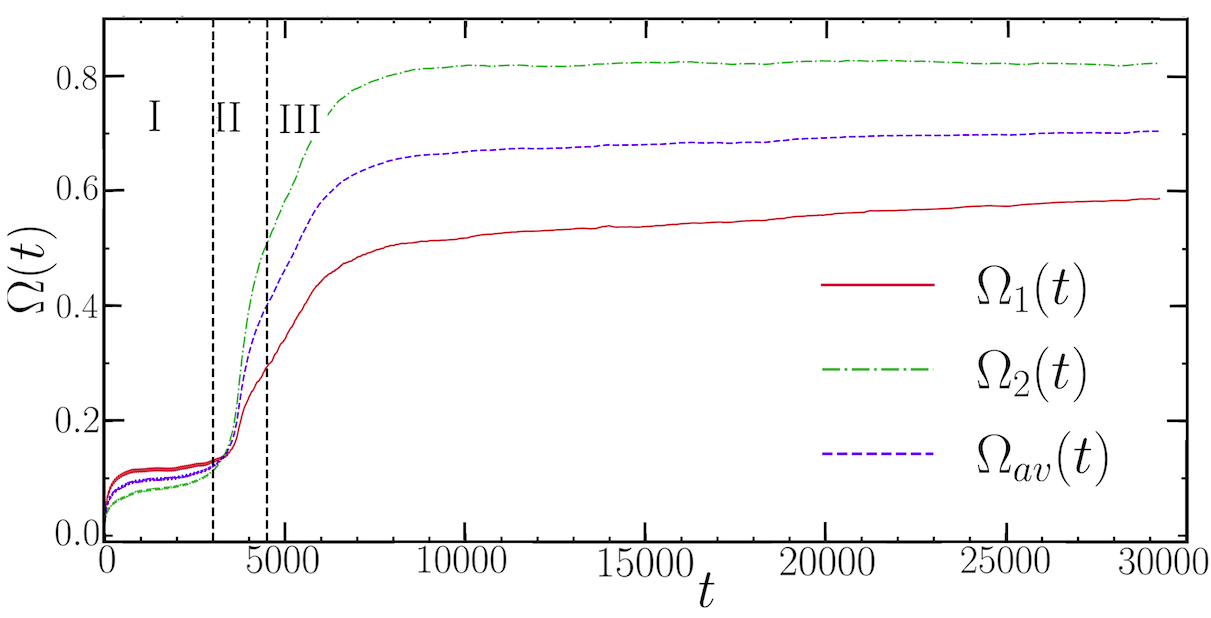}
\end{center}
\caption{(Color on line)  The time evolutions of the classical rotational velocity, $\Omega_i(t)$,
given by \eqref{rot-cl}, for both components, $^{85}$Rb (solid-red line)  and $^{133}$Cs (dot-dashed-green line), 
together with the corresponding averaged result, $\Omega_{av}(t)\equiv[\Omega_1(t)+\Omega_2(t)]/2$
(dashed-blue line). The vertical lines are approximately separating three time intervals in the evolution: 
(I) shape deformations; (II) turbulent regime, with nucleation of vortices; and (III) with the vortex lattices 
being estabilized. All quantities are dimensionless, with time units in ${\omega_\rho^{-1}}$.
}
\label{f04}
\end{figure}

The time evolutions obtained for the classical rotational velocities $\Omega_i(t)$ are shown in Fig.~\ref{f04},
obtained from the expected value ratios between the induced angular momentum $\langle L_z\rangle_i$
and the corresponding momentum of inertia $\langle I\rangle_i $, as given by Eq.~\eqref{rot-cl}.
 The rotating condensates always reflect the angular momentum, which is a conserved quantity, at the equilibrium state. 
 When the angular  momentum of the system increases,  the rotational velocity also increases, correspondingly. 
 We can clearly observe three time intervals in the evolution of the rotational velocity, given by Eq.~\eqref{rot-cl},
 as indicated inside the figure. The initial interval (I) corresponds to the period in which we have the shape 
deformations of the condensates. The nucleation of vortices occurs in the region identified by (II) in the figure, which 
corresponds to the increasing of the velocities observed in Fig.~\ref{f03}(a). In the region (III), we have the
vortex lattice formations, with $\Omega_i(t)$ converging to almost constant values.

The overall time evolution of the expected values will provide the corresponding time-dependent 
defined classical rotational frequency for each component of the mixture, given by a red-solid line for the lighter
species ($^{85}$Rb) and a green-dot-dashed line for the heavier species  ($^{133}$Cs). As shown in Fig.~\ref{f04}, 
the frequencies grow faster for the lighter component than for the heavier one in the initial time interval, when 
the coupled system is still in the time interval given by region I, before the nucleation of vortices. In region II, we 
notice the transition with vortices nucleation at the surface, when the frequencies start to grow faster for the heavier 
element. The stabilization of the frequencies are verified in the long-time interval given by region III, with
the emergence of stable vortex patterns. 

\begin{figure}[tbp]
\begin{center}
\includegraphics[width=0.45\textwidth]{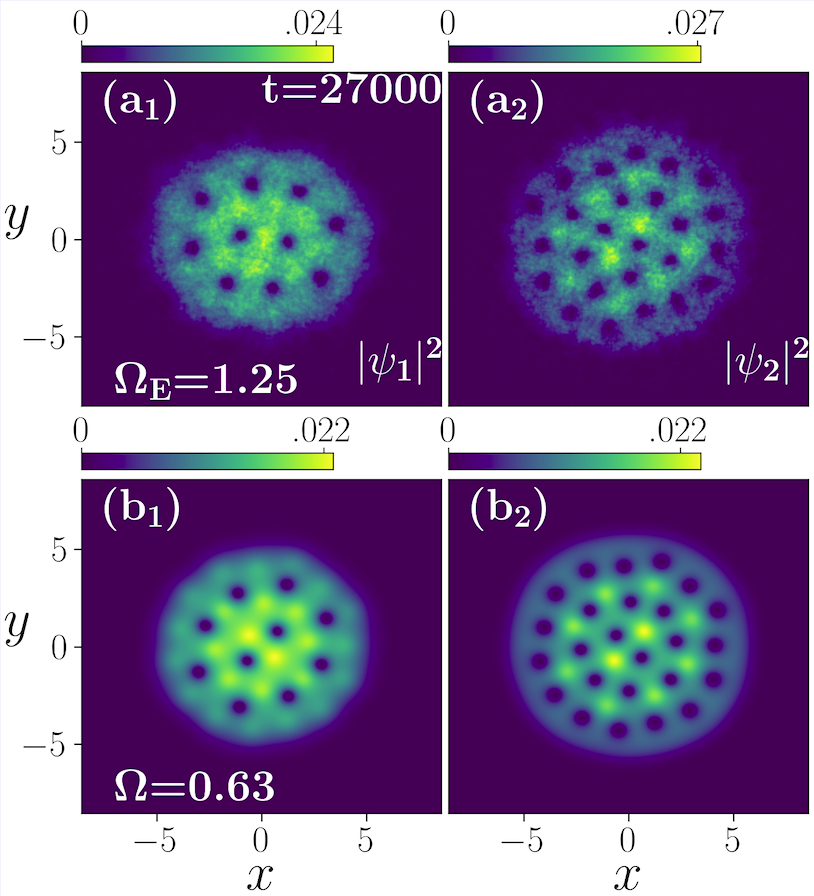}
\end{center}\vspace{-.5cm}
\caption{(Color on line) The vortex-pattern solutions, obtained at $t=27000$ with the stirring
potential  [(a$_i$) = (d$_i$) of Fig.~\ref{f01}], are compared with the ground-state solutions of
\eqref{2d-2c}, with $\Omega_0=0.63$ and $V_s=0$, shown in panels (b$_i$), for the
miscible configuration $a_{12}/a_{11}=0.5$.
 The density levels are indicated at the top of each panel.
All quantities are dimensionless, with $l_\rho$ and $\omega_\rho$ being the respective length and 
frequency units.}
\label{f05}
\end{figure}

In order to simulate the ground-state solution with a time-independent single rotational frequency, $\Omega_0$, 
we consider that this frequency $\Omega_0$ will correspond to a time averaging of both rotational frequencies
$\Omega_i(t)$, in which we assume for the total averaging time the interval from zero up to the point in which the frequency becomes almost constant, such that $\Omega_i(t\ge T)\approx\Omega_i(T)$. 
With this prescription, we obtain
\begin{equation}\label{average}
\Omega_0\equiv\frac{1}{T}\int_0^T\Omega_{av}(t)dt = \frac{1}{2T}\int_0^T[\Omega_1(t)+\Omega_2(t)]dt.
\end{equation}
From the results shown in Fig.~\ref{f04}, in which $T\sim 30000$ and the saturated averaged frequency is 
$\Omega_{av}(T)\sim 0.72$, the above prescription will give us $\Omega_0\sim 0.63$.
As shown in Fig.~\ref{f05}, this prescription  is providing an approximate good value for the classical
rotational frequency, when using in Eq.~\eqref{2d-2c}, without the stirring potential.

In Fig.~\ref{f05} we provide a comparison between the results for the coupled densities, obtained with the 
stirring potential [panels (a$_i$)] and without it [panels (b$_i$)], by using the prescription (\ref{average}). 
With the stirring potential, we assume $\Omega_0=0$ in Eq.~\eqref{2d-2c}, and consider the 
laser stirring potential  \eqref{2Dtrap}, with $\Omega_E=1.25$ and $\epsilon=0.025$. The corresponding 
stable results, presented in panels (a$_1$) and (a$_2$) are the ones previously shown in the panels (d$_1$) 
and (d$_2$) of Fig.~\ref{f01}, respectively, for $t=27000$.
Without the stirring potential ($\epsilon=0$), we obtain the complete ground-state solutions of the coupled 
densities, by considering the frequency $\Omega_0$ in  Eq.~\eqref{2d-2c} as given by the time averaging
frequency \eqref{average}, which will give us approximately $\Omega_0=0.63$ in the present case.
The results verified in the panels (b$_1$) and (b$_2$) of Fig.~\ref{f05}, computed  directly for the 
ground-state solutions, as obtained for the ground-state solutions are not affected by sound-wave propagations.

\begin{figure}[tbp]
\begin{center}
\includegraphics[width=0.4\textwidth]{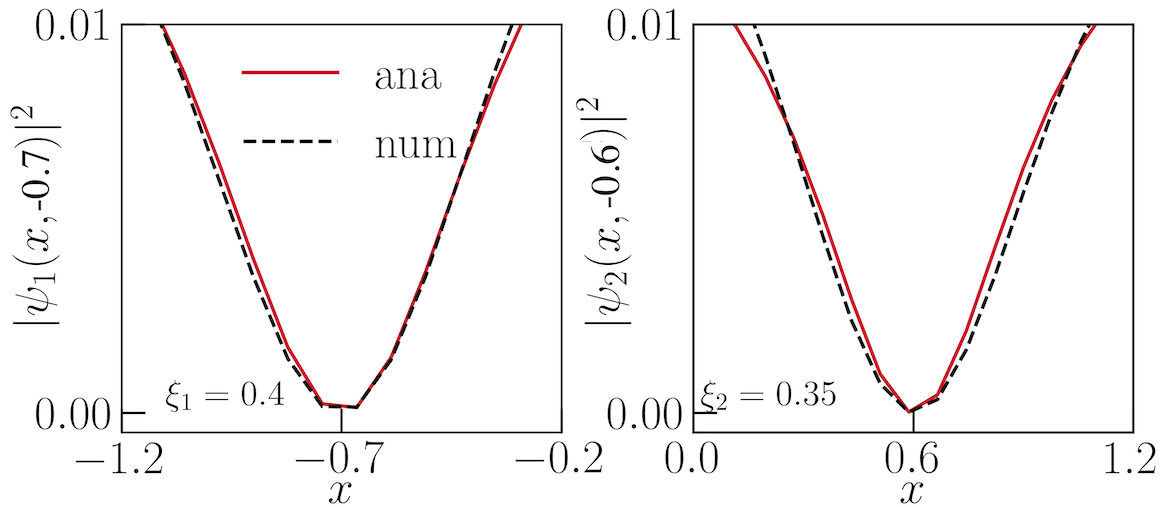}
\end{center}\vspace{-.5cm}
\caption{One-dimensional density views of the cores of two vortices shown in Figs.~\ref{f05}(b$_1$) and (b$_2$),
respectively centered at $(x,y)=(-0.7,-0.7)$ and $(x,y)=(0.6,-0.6)$, obtained for the two components. 
The numerical solutions for the vortex-core sizes (dashed lines) are being compared with the 
corresponding analytical ones,  fitted by 
 $\eta_i(x)=\eta_{0,i}[1+\xi_i^2/(x-x_c)^2]^{-1}$, where the given rotational states healing lengths,
 $\xi_1=0.4$ and $\xi_2=0.35$ are obtained by the analytical parameterization, with 
$\eta_{0,1}=0.019$ and $\eta_{0,2}=0.015$ giving the density maxima near the vortex-cores.  }
\label{f06}
\end{figure}
 The healing length is directly related to the vortex size in a superfluid. 
To estimate the two species healing lengths from our numerical solutions shown 
in the panels (b$_1$) and  (b$_2$) of Fig.\ref{f05}, we display two corresponding
panels in Fig.\ref{f06}. So, two vortex cores near the center of the condensates are selected, 
in order to estimate the corresponding vortex-core sizes. 
They are plotted in the panels (a) and (b) of Fig.~\ref{f06}, by fixing the vortex center position
$y_c$, allowing the $x$ position be varied near the vortex center $x_c$.
For $^{85}$Rb, we consider the vortex that appears in Fig.~\ref{f05}(b$_1$) centered at 
$(x_c,y_c)\approx (-0.7,-0.7)$, which is shown in Fig.~\ref{f06}(a).
For $^{133}$Cs, the selected vortex is in Fig.~\ref{f05}(b$_2$) centered at 
$(x_c,y_c)\approx (0.6,-0.6)$,  shown in Fig.~\ref{f06}(b). 
So,  the numerical solutions for the vortex-core sizes (dashed lines) are being compared with 
the corresponding analytical ones, fitted by 
$\eta_i(x)=\eta_{0,i}\left[1+{\xi_i^2/(x-x_c)^2}\right]^{-1}$, where $\xi_i$ 
are fitting parameters for the rotational state  healing lengths, with $\eta_{0,i}$ the density 
maxima near the vortex cores.

\begin{figure}[tbp]
\begin{center}
\includegraphics[width=0.45\textwidth]{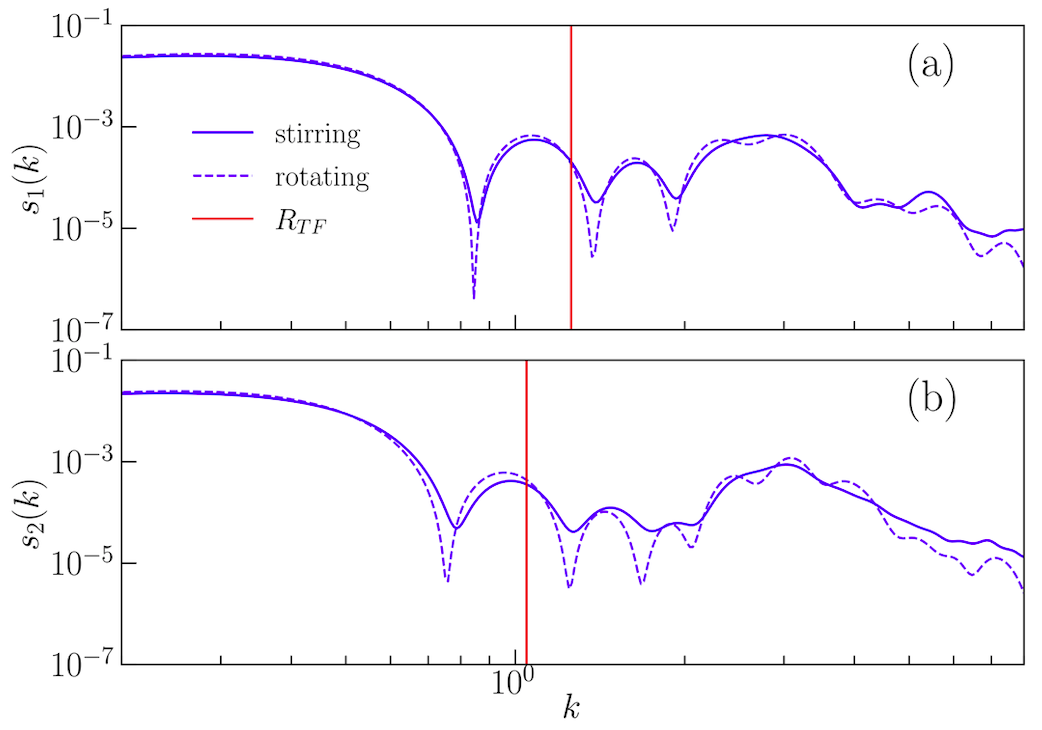}  
\end{center}
\vspace{-.5cm}
\caption{(Color on line) 
The structure factor solutions $s_i(k)$ [panel (a) for $^{85}$Rb, with panel (b) for $^{133}$Cs], obtained
with the stirring potential (blue-dashed lines) and by rotating without the stirring (blue-solid lines).  
Plotted as functions of $k$, the minima correspond to the lattice orders. 
The vertical red lines in both panels are indicating the respective Thomas-Fermi (TF) inverse radius 
positions. All quantities are dimensionless.}
\label{f07}
\end{figure}

In Fig.~\ref{f07}, the real time and ground-state solutions for the structure factors are plotted as functions 
of the wave number $k$, by applying the Eq.~\eqref{sf}. As verified, the minima for the oscillations, which will 
correspond to the lattice orders, occur at about the same positions.

{
\subsection{Binary $^{85}$Rb$-^{87}$Rb mixture}
For the case of two isotopes of the same species we assume the binary mixture $^{85}$Rb$-^{87}$Rb.
The main purpose, by repeating the same kind of calculations previously done for the
$^{85}$Rb$-^{133}$Cs, is to verify clear mass-imbalance effects in our results. Particularly, we will try 
to point out the system which is more favorable in the production of vortex patterns using with a 
stirring potential.
For that, we follow the previous same parameterizations for the trap and particle interactions,
considering a complete miscible configuration, with $\delta = a_{12}/a_{11}=0.5$ (where, $a_{22}=a_{11}=60a_0$).

To time evolution of the densities can be verified by the four set of panels shown in Fig.~\ref{f08}, for the two species, 
with the upper row being for the $^{85}$Rb and the lower row for the $^{87}$Rb. As expected, due to the small
mass difference between the components, the dynamical behavior observed for both species is similar along the
time evolution, till the vortex-patterns becomes stable. 
In comparison with Fig.~\ref{f01}, we should observe that it is shorter the time to reach the condition in which
occurs the surface nucleated vortices move from surface to the inner part of the condensates.
This behavior can better be verified as considering the results obtained for the other observables, in
correspondence with the previous stronger mass-imbalanced case.
For that, we are presenting the time-evolution results for the energies in Fig.~\ref{f09},  
currents and torques in Fig.~\ref{f10}, with the corresponding classical rotational velocities shown
in Fig.~\ref{f11}.

\begin{figure*}[htbp]
\begin{center}
\includegraphics[width=0.9\textwidth]{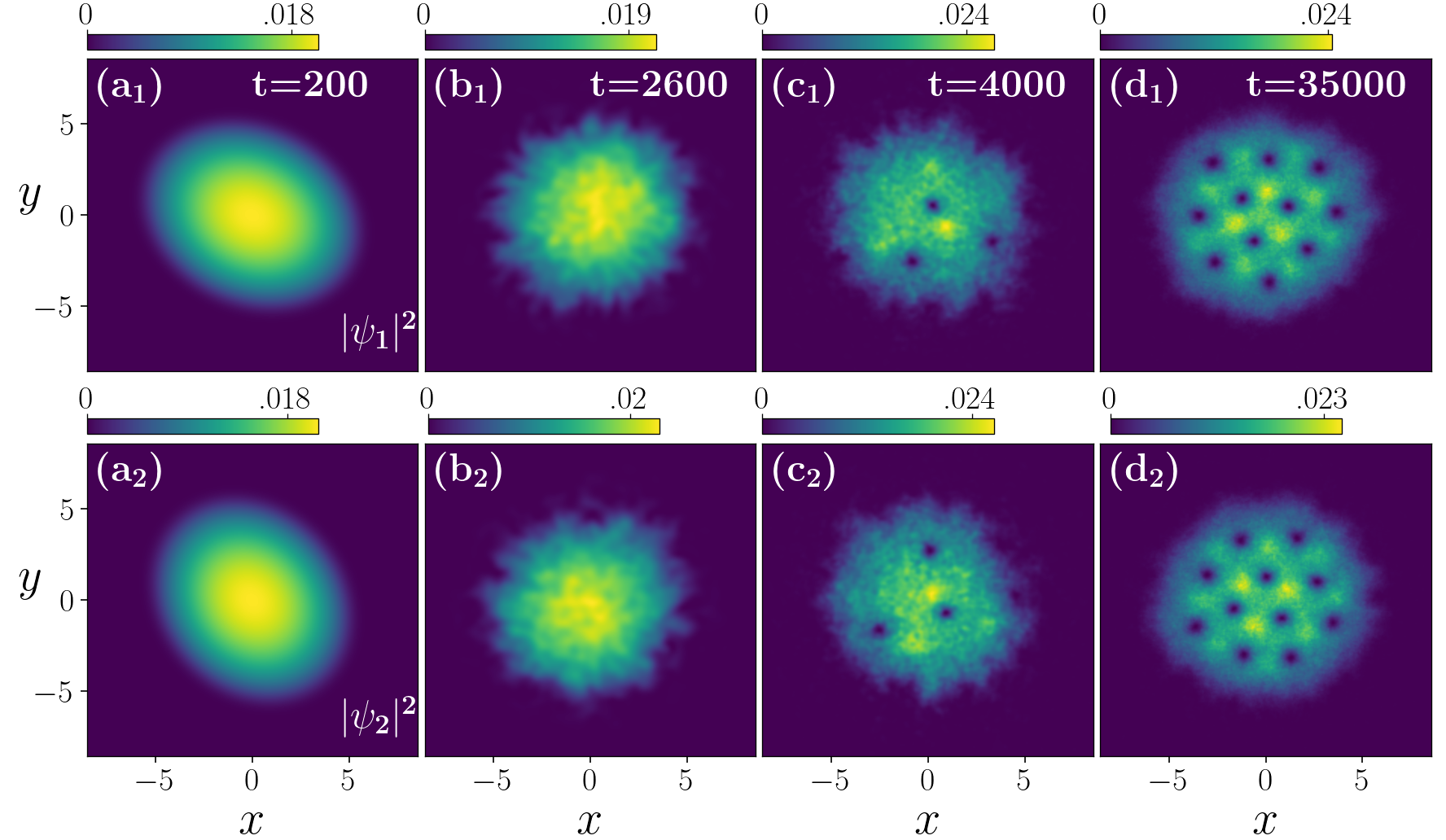}
\end{center}
\vspace{-.5cm}
\caption{(Color on line) Time evolution of the densities for both condensed species of mixture $^{85}$Rb-$^{87}$Rb. 
The upper panels are for the $^{85}$Rb (component 1), with the lower panels for $^{87}$Rb (component 2). 
As in the case of Fig.~\ref{f01}, by starting with a ground-state solution, the system is evolved in real time with 
$\Omega_E=1.25$ with $\epsilon=0.025$.  The density levels are indicated at the top of each panel.
All quantities are dimensionless, with $l_\rho$ and $\omega_\rho$ being the respective length and 
frequency units. 
}
\label{f08}
\end{figure*} 
\begin{figure}[htbp]
\begin{center}
\includegraphics[width=.8\columnwidth]{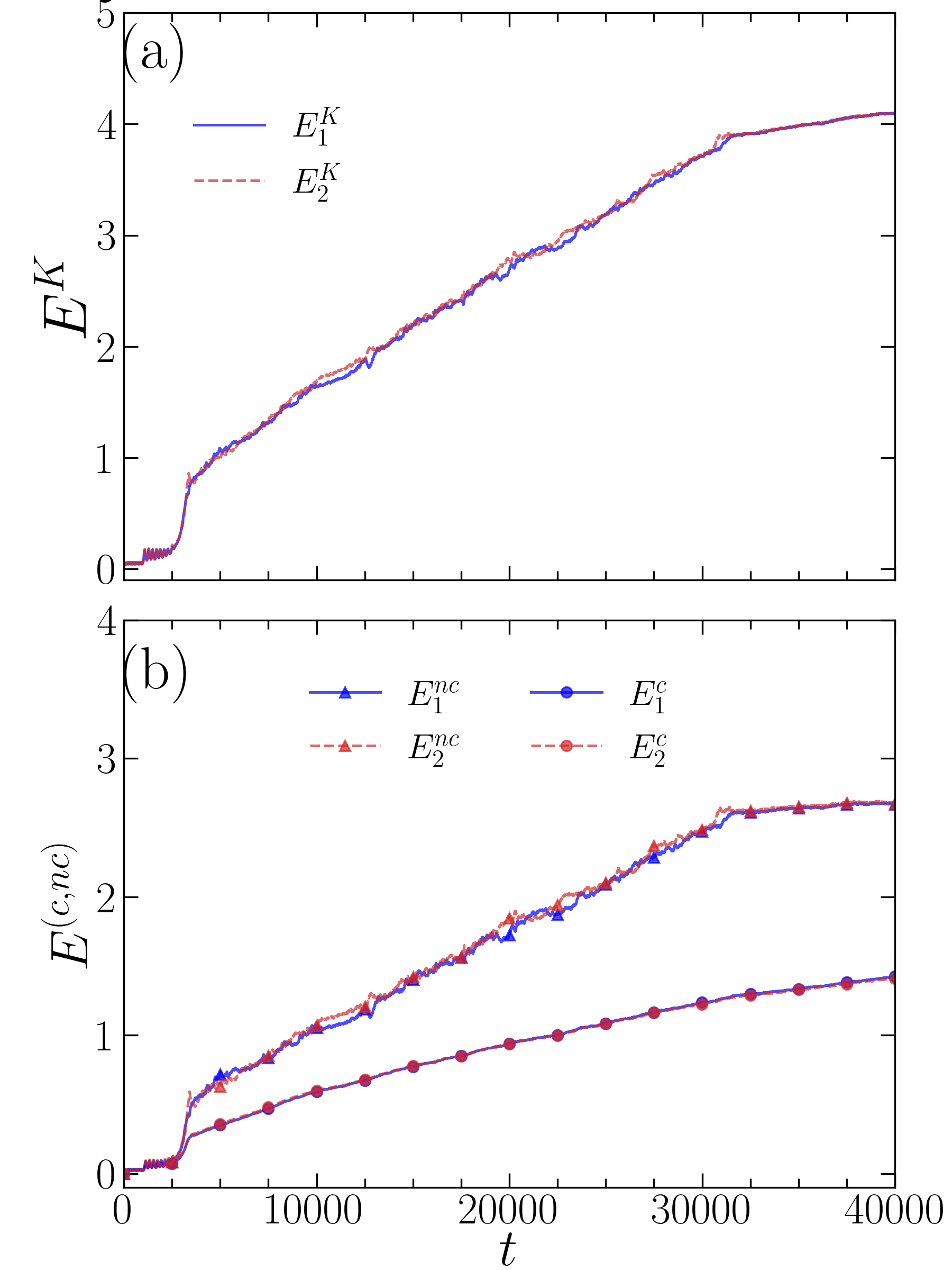}
\end{center}\vspace{-.5cm}
\caption{(Color on line) 
For both components of the mixture, $^{85}$Rb ($i=1$) and $^{87}$Rb ($i=2$), the panel (a) shows the almost 
complete overlap of the total kinetic energies (solid-blue for $E_1^K$, with dashed-red for $E_2^K$).
The corresponding compressible (with bullets) and incompressible (with triangles) parts of the kinetic 
energies are shown in  panel (b).
All quantities are dimensionless, considering the defined units.
}
\label{f09}
\end{figure}
\begin{figure}[htbp]
\begin{center}
\includegraphics[width=0.45\textwidth]{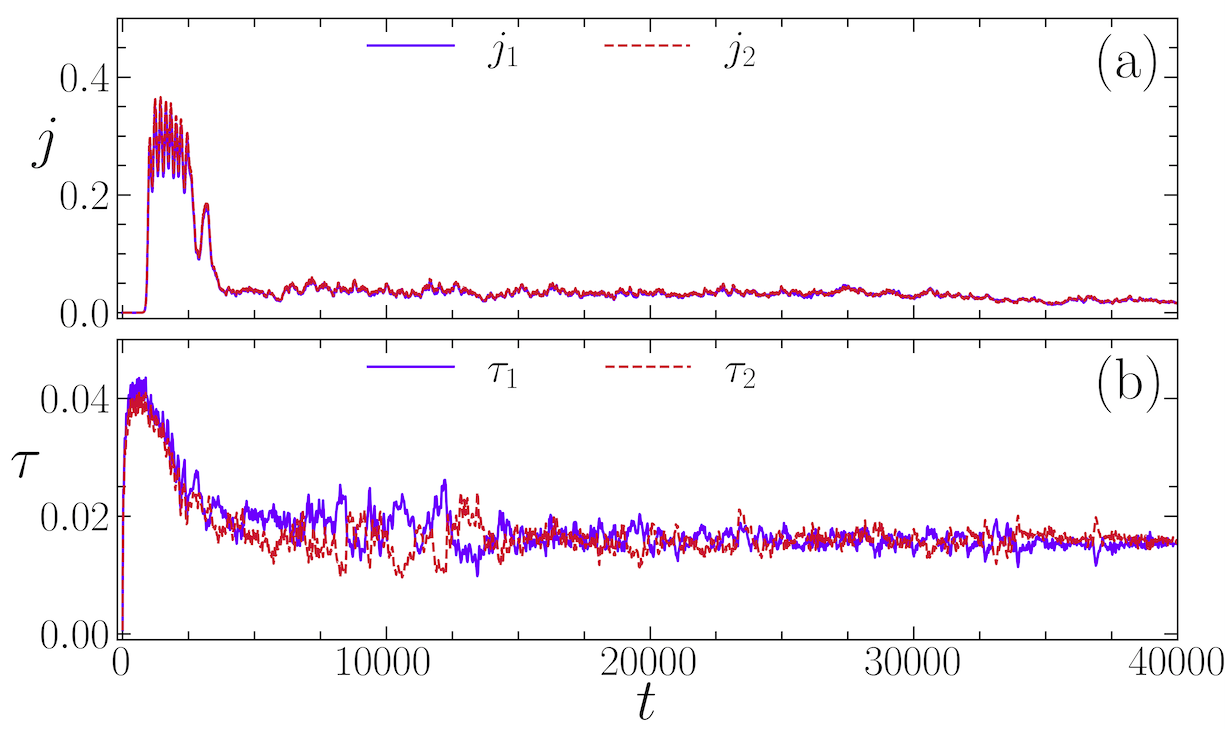}
\end{center}\vspace{-.5cm}
\caption{(Color on line) The current densities  
[panel (a)] with the torques  
[panel (b)], respectively given by Eqs.~\eqref{current} and \eqref{torque} for both components, 
are shown for the  $^{85}$Rb-$^{87}$Rb mixture, during the dynamical process that occurs till the 
vortex generation.
}
\label{f10}
\end{figure}
\begin{figure}[htbp] 
\begin{center}
\includegraphics[width=8.5cm]{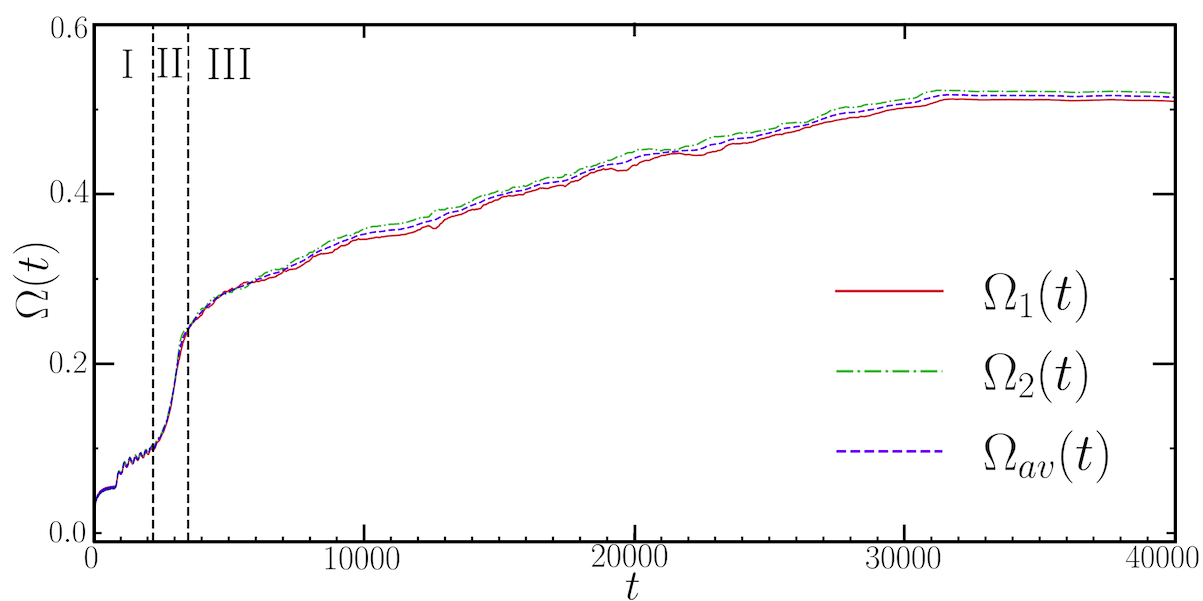}
\end{center}\vspace{-.5cm}
\caption{(Color on line)  This figure shows the time evolution of the classical rotational velocity 
$\Omega_i(t)$, obtained for both components, $^{85}$Rb (solid-red curve)  and $^{87}$Rb (dot-dashed-green curve), 
together with the corresponding averaged result, given by  $\Omega_c(t)=[\Omega_1(t)+\Omega_2(t)]/2$
(dashed-blue curve), given by Eq.~\eqref{rot-cl}. The vertical lines are indicating the separation of three time
intervals in the evolution: (I) before the nucleation of vortices, when we have just the shape deformations;
(II) transition period with the nucleation of vortices at the surface; and (III) with the vortex lattices being 
formed in the binary condensate. 
}
\label{f11}
\end{figure}
\begin{figure}[tbp]
\begin{center}
\includegraphics[width=0.47\textwidth]{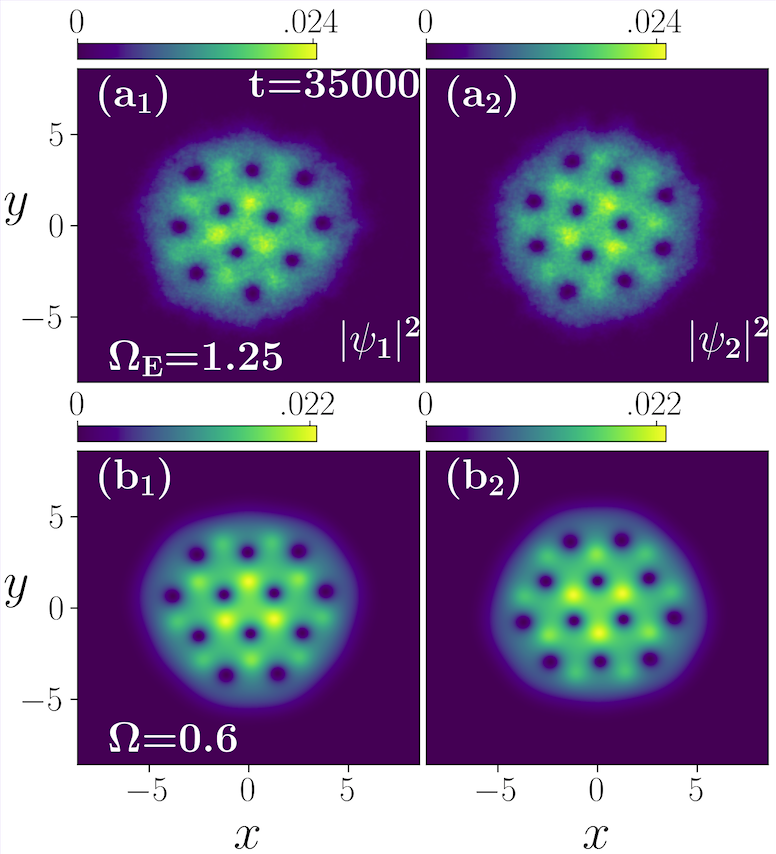}
\end{center}\vspace{-.5cm}
\caption{(Color on line) The ground-state stable density solutions for the coupled
system $^{85}$Rb (species 1) and $^{87}$Rb (species 2), with $\Omega_0=0.6$ and
without the stirring potential, are shown in the above set of panels (b$_1$) and (b$_2$). 
They are compared with the respective two-species results
obtained with the stirring potential ($\Omega_E=1.25$ and $\epsilon=0.025$),
at $t=35000$, shown in (a$_1$) and (a$_2$).
In both the cases, the couple system is in a mixed configuration with $a_{12}/a_{11}=0.5$.
 The density levels are indicated at the top of each panel. All quantities are dimensionless, 
 with $l_\rho$ and $\omega_\rho$ being the respective length and frequency units.}
\label{f12}
\end{figure} 

In this case, with both species having about the same mass, the final number of vortices verified for both 
components is the same, as shown in Fig.~\ref{f08}.
As the respective results are very close, with total energies much higher than the kinetic energies,
a single panel is presented for the energies in Fig.~\ref{f09}, where the total energies are in the upper region with
the compressible and incompressible parts of the kinetic energies in the lower part of the panel.
These results, together with the results shown for the current and torques, are also quite in contrast with the
corresponding ones verified for the stronger mass-imbalanced case, as the process of vortex nucleations and
movement to the inner part of the condensate become much faster. However, we should notice that the time 
to stabilize the final vortex patterns is much longer, as one can verify in particular from the results obtained
 the classical rotational frequencies. By comparing Fig.~\ref{f11} with Fig.~\ref{f04}, we noticed that the 
 increasing in the averaged frequencies have already saturated for $t\sim 30000$ when the case 
 $^{85}$Rb$-^{133}$Cs; but, when considering  $^{85}$Rb$-^{87}$Rb, we need to go to a much longer
 time.

\section{Incompressible kinetic energy spectra}\label{sec4}
{
In this section, we consider the velocity power spectral densities in the $k$ space, as given by Eqs.~(\ref{spec1}) 
and (\ref{spec}), for the analysis of the incompressible kinetic energy spectra of the two components, 
which is the more appropriate part of the kinetic energy concerned the fluidity characteristics in the vortex production.
 The kinetic energy spectral methods for analyzing turbulent flows in symmetry-breaking
quantum fluids with in the Gross-Pitaevskii limit can be found in Ref.~\cite{2021Bradley}.
The incompressible $E^{(nc)}$ spectrum is expected to behave as $k^{-3}$ when vortex configurations are well established 
inside the condensate. In the time evolution regime, before reaching the vortex-pattern configuration, the spectrum is 
expected to approach the $k^{-5/3}$ Kolmogorov behavior, characterizing the turbulent regime of a fluid.
In order to enhance the effect of mass-differences in this power-law dynamical behavior study, both mass-imbalanced 
coupled systems that we are considering,  $^{85}$Rb-$^{133}$Cs and  $^{85}$Rb-$^{87}$Rb, will be investigated 
comparatively along the time-evolution of the mixtures.

\begin{figure*}[!t]
    \begin{center}
    \includegraphics[width=0.99\textwidth]{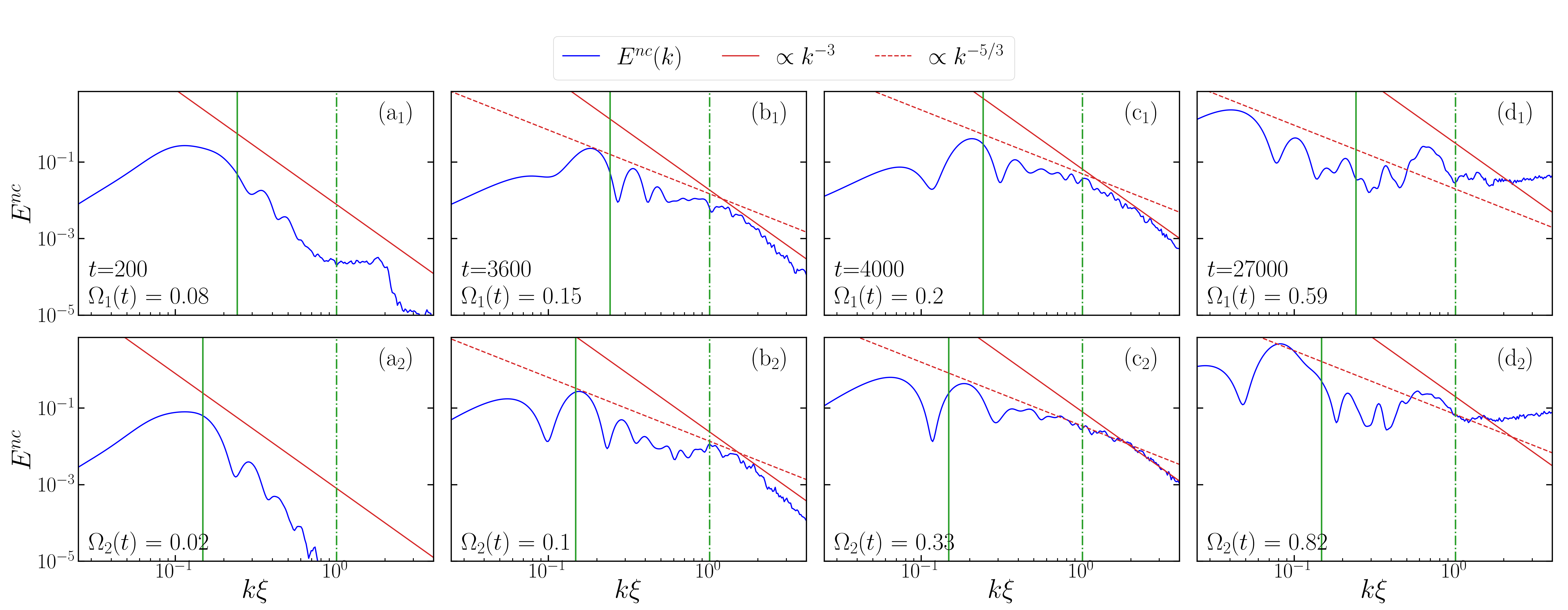}
    \caption{(color online) Corresponding to four instant times shown in Fig.~\ref{f01}, for the coupled
    system $^{85}$Rb-$^{133}$Cs, in this figure we present four sets of log-scaled plots for the 
    incompressible kinetic energy spectra,$E^{nc}\equiv E^{nc}(k,t)$, versus $k\xi$, where $k$ is the wave number 
    and $\xi$ the healing length.     The spectrum is calculated in rotating frame, with the upper panels (a$_1$-d$_1$) 
    for the first component $^{85}$Rb,  and bottom panels (a$_2$-d$_2$) for $^{133}$Cs. 
    The red-solid lines indicate the $k^{-3}$ behavior; with the red-dashed
    lines being for the $k^{-5/3}$ behavior. The solid-green vertical lines indicate the inverse Thomas-Fermi (TF) radial    
positions (given by $k\xi=2\pi\xi/R_{TF}$), with the dot-dashed-green lines, at $k\xi=1$, to estimate the approximate 
positions at which the power-law behavior is changing.  All quantities are dimensionless, with the length, time and energy units 
defined in the text. }
    \label{f13}
    \end{center}
\end{figure*}
\begin{figure*}[!t]
    \begin{center}
    \includegraphics[width=0.99\textwidth]{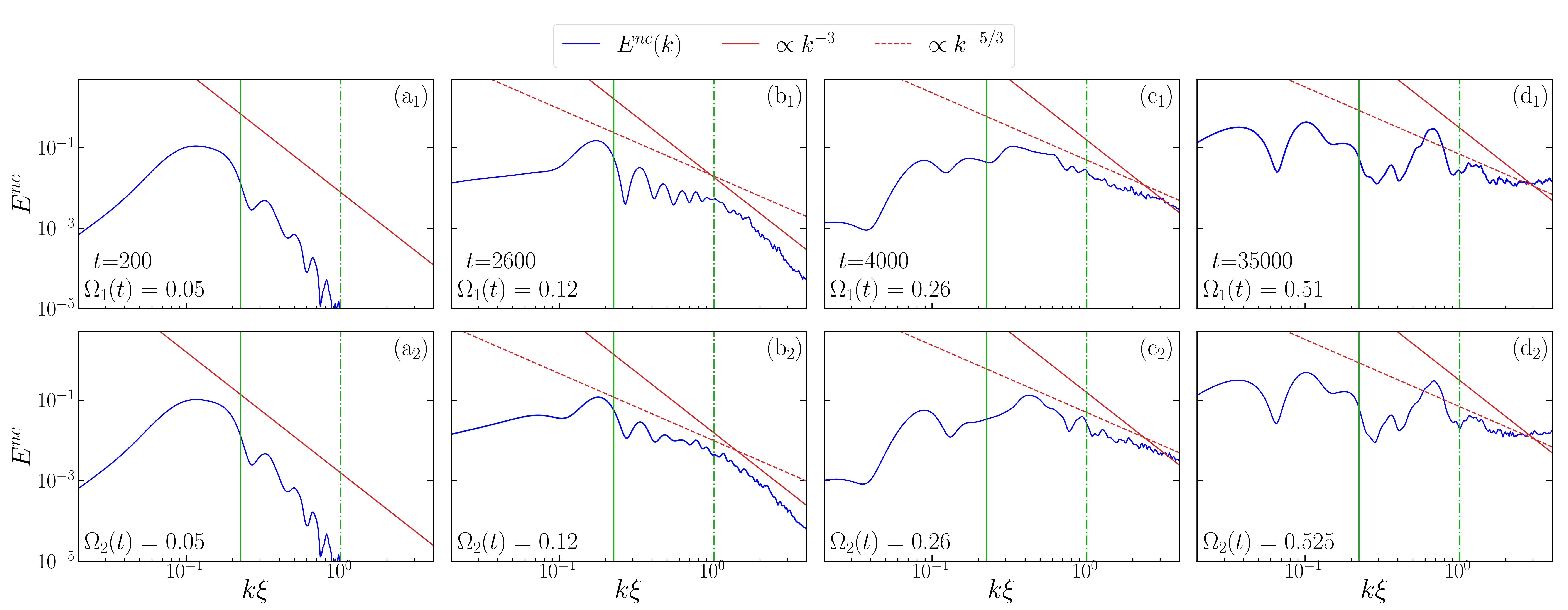}
    \caption{(color online) Here, it is shown the incompressible kinetic energy spectra, $E^{nc}\equiv E^{nc}(k,t)$, for the the 
    $^{87}$Rb-$^{85}$Rb mixture, with the  selected times following the panels shown in Fig.~\ref{f08}.    
     The spectrum is calculated in rotating frame, with the upper panels (a$_1$-d$_1$) for the first component $^{85}$Rb, 
     and bottom panels (a$_2$-d$_2$) for $^{87}$Rb.  The line conventions and units are as detailed in the caption of Fig.~\ref{f13}.
}
    \label{f14}
    \end{center}
\end{figure*}

\begin{figure}[!t]
    \begin{center}
    \includegraphics[width=0.95\columnwidth]{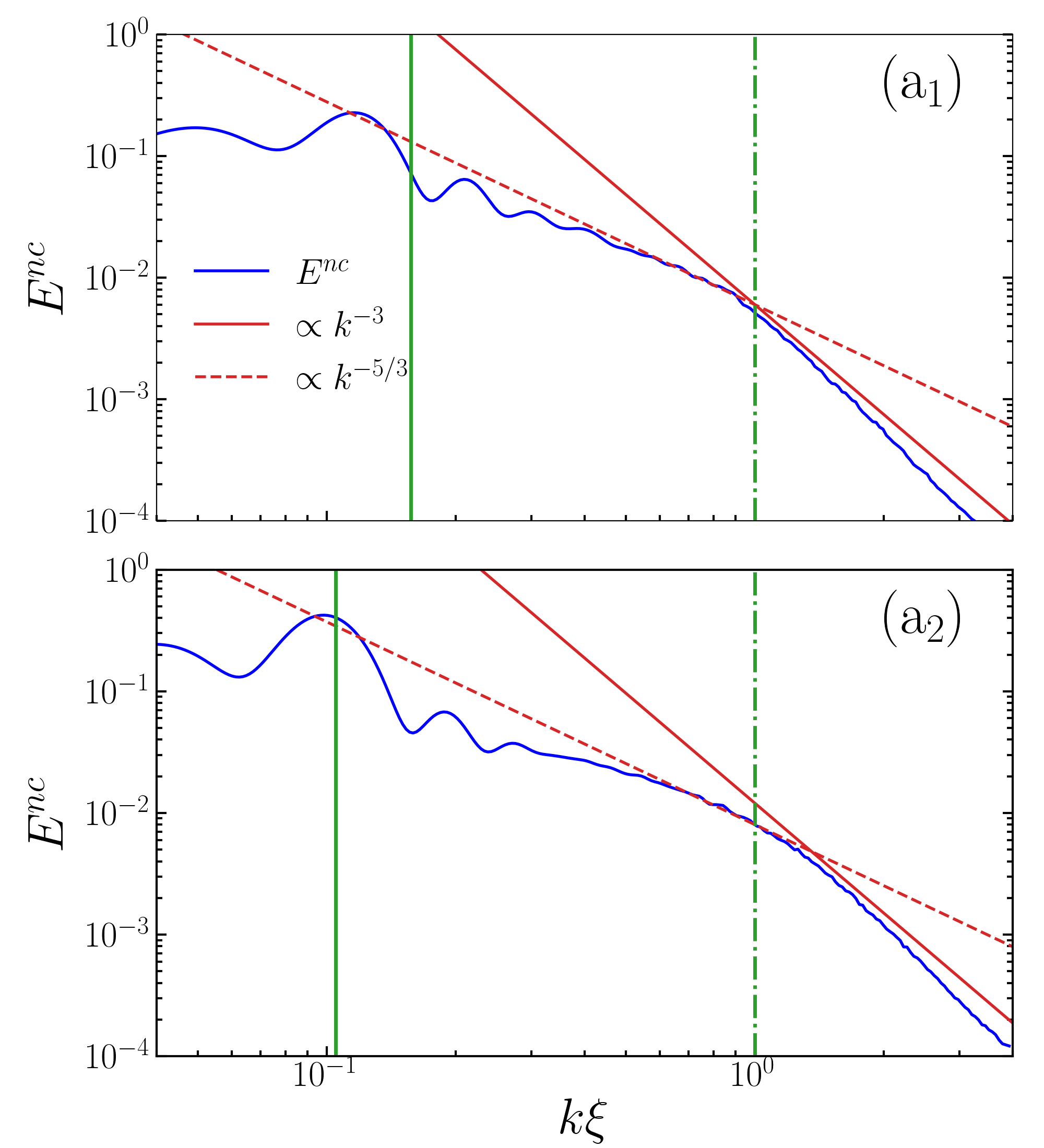}
    \caption{(color online) Incompressible kinetic energy spectra, $E^{nc}\equiv E^{nc}(k,t)$, for 
     the $^{85}$Rb-$^{133}$Cs mixture, obtained 
by averaging over 50 samples in the turbulent time-interval regime II defined in Fig.~\ref{f04}, confirming the
Kolmogorov $k^{-5/3}$ power law behavior in this interval (red-dashed lines), being modified to  
the $k^{-3}$ behavior when the vortex patterns are established. 
The line conventions and units are as detailed in the caption of Fig.~\ref{f13}.}
    \label{f15}
    \end{center}
\end{figure}

\begin{figure}[!t]
    \begin{center}
    \includegraphics[width=1.\columnwidth]{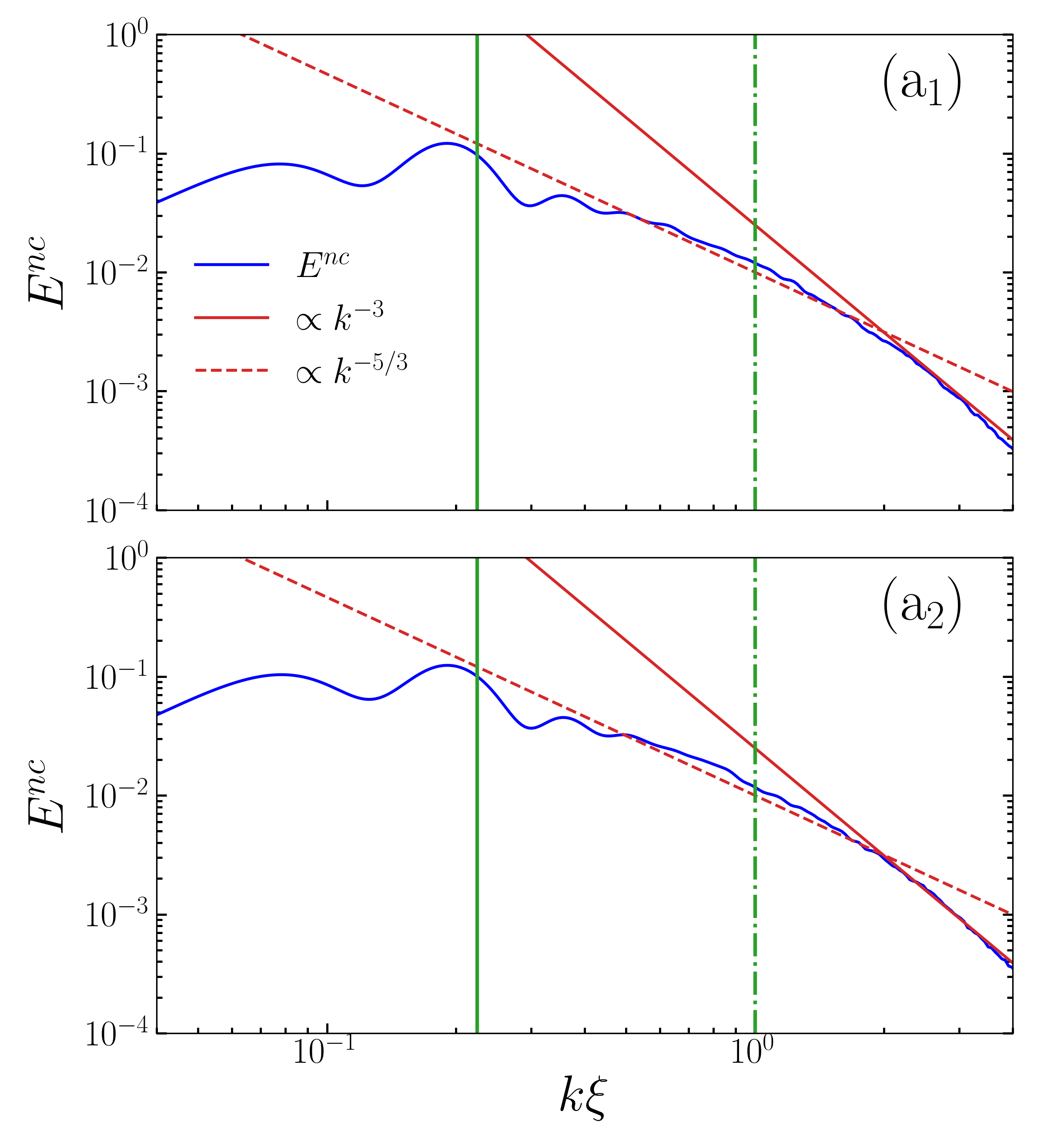}
    \caption{(color online) Incompressible kinetic energy spectra, $E^{nc}\equiv E^{nc}(k,t)$, for
     the $^{85}$Rb-$^{87}$Rb mixture, by averaging over 50 samples in the turbulent regime II, indicated in 
     Fig.~\ref{f11}. The behavior is similar as the one observed in Fig.~\ref{f15} for the stronger 
     mass-imbalanced case.  As verified by comparing with the $^{85}$Rb-$^{133}$Cs mixture, the 
     change in the power-law behavior occurs for larger values of $k\xi$. 
     The line conventions and units are as detailed in the caption of Fig.~\ref{f13}.}
    \label{f16}
    \end{center}
\end{figure}
Therefore, in the next, both mass-imbalanced coupled systems, $^{85}$Rb-$^{133}$Cs and  $^{85}$Rb-$^{87}$Rb, have 
their incompressible kinetic energy spectra, $E^{(nc)}(k,t)$, presented in Figs.~\ref{f13} to \ref{f16}. The results are
displayed as functions of the wave-number $k$ multiplied by the corresponding healing lengths $\xi$. 
For both the cases, in the selected set of coupled panels we are considering time instants, with the corresponding classical rotational
 frequencies, which are representative of the time intervals before and after the nucleation of the vortices.  

\subsection{$^{85}$Rb-$^{133}$Cs kinetic energy spectra}
With Fig.~\ref{f13},  we first show the kinetic energy spectra for the case in which a stronger 
mass imbalance exists between the species. 
By considering the three regions defined in Fig.~\ref{f04}, four representative times of the evolution were
chosen, with the upper panels for the lighter species ($^{85}$Rb) and lower panels for heavier one
($^{133}$Cs).

In the panels  (a$_i$), for $t=200$, the two-component spectra correspond to the shape 
deformation interval, when we noticed that the rotation frequency is much slower for the heavier component. 
Since, there is no vortex formation at this stage, analyzing spectrum in this regime does not provide 
any useful information about the power-law behavior. 
The panels (b$_i$), at $t=3600$, refer to the turbulent interval at which vortices start to be formed at the surface;
with the panels (c$_i$), at $t=4000$, shown the spectra when the vortices are already entering the inner part of 
the coupled condensate.  The vortices appearing in the surface of the condensate are already following 
the $k^{-3}$ behavior in the ultraviolet regime ($k\xi\gg 1$).
Significantly, a quantum fluid is more turbulent at this stage due to large number of vortex generation at the surface. 
The quantum fluid looks more turbulent until the torque, experienced by the condensates due to laser stirring, arrives at 
equilibrium.  
The turbulent behavior is clearly characterized by the $k^{-5/3}$ power-law behavior as shown in (b$_i$)-(c$_i$). 
Along the process, the coupled condensates have a continuous energy injection due to the laser stirring potential.
We use the constant stirring potential, 
with the coupled system reaching the equilibrium, after the maximum transfer of energy, due to the conservation of angular momentum.
Finally, the panels (d$_i$), at $t=27000$, are showing the spectra when the patterns of vortices became already stable,
with the rotation frequency of the heavier component becoming much faster than the lighter one. 
In these panels (d$_i$) results, one can observe that the sound waves start hassling the vortex cores,  confirmed by the deviation 
from the $k^{-3}$ behavior.  Here, one should not expect any turbulent behavior, with the system already 
reaching the equilibrium.

For these four time instants, the corresponding density distributions in the configuration space can be seen 
in Fig.~\ref{f01}. The straight lines (solid and dashed), inside the panels,  are indicated the two main behaviors for 
the kinetic energy spectra, which are the $k^{-3}$ corresponding to the ultraviolet regime, when the vortex 
patterns are formed; and the $k^{-5/3}$ behavior related to the turbulent intermediate regime.

The velocity power spectrum is  dominated by the rigid-body rotation spectrum in the infrared regime. 
In this work, we remove the large infrared contribution by calculating the spectrum in the rotating frame,
as the amplitude of the infrared signal is greatly reduced  in such frame.
The incompressible kinetic energy spectra, corresponding to the vortex configurations 
shown in Fig.~\ref{f01}, are shown over wavenumber in the units of healing 
length $k\xi$ in log scale. Similarly as in Ref.~\cite{2012Bradley}, also considering our present results
of coupled systems, the spectra are analyzed for the infrared ($k\xi\ll2\pi$) and ultraviolet 
($k\xi\gg2\pi$) regimes. 
In the ultraviolet regime, the incompressible kinetic energy spectrum has a 
universal $k^{-3}$ scaling behavior that arises from the vortex core structure. The infrared regime 
arises purely from the configuration of the vortices.  The incompressible spectrum becomes more 
important when vortices enter into the system.  After the symmetry is broken, we observe in the
panels (b$_i$) and  (c$_i$) turbulent behaviors being created for both components 
as the  vortices start to appear at the surface of the condensate. 
In this infrared regime,  
the incompressible spectrum goes in agreement with the Kolmogorov $k^{-5/3}$ power law.
However, this  $k^{-5/3}$ power law feature is kept till $k\xi\sim 2\pi$, changing to $k^{-3}$
at higher momenta. These behaviors vanishes, after the vortices have entered in the inner part 
of the densities, with the coupled system relaxing to the crystallization of lattice patterns.
In order to better characterize this regime, in Fig.~\ref{f15} we present 
the incompressible kinetic energy spectra of $^{85}$Rb-$^{133}$Cs mixture, which were 
obtained by averaging over 50 samples in the turbulent time-interval regime II defined in Fig.~\ref{f04}, 
which confirms the Kolmogorov $k^{-5/3}$ power law behavior in the turbulent regime II, for
$k\xi\lesssim 1$, changing to $k^{-3}$ for larger momenta.

\subsection{$^{85}$Rb-$^{87}$Rb kinetic energy spectra}
Correspondingly,  in Fig.~\ref{f14} we have our results for the spectra when considering the   $^{85}$Rb-$^{87}$Rb
mixture. In this case, following the same order as before, we have the initial shape deformation interval being represented
in  (a$_i$), at $t=200$. The other time instants are for the time interval when the nucleation of vortices start 
[(b$_i$), at $t=2600$], entering the inner part [(c$_i$), at $t=4000$], and when the vortex-patterns become stable 
[(d$_i$), at $t=35000$].  For this mixture, the corresponding density distributions are given in Fig.~\ref{f08}.

The results for the incompressible kinetic energy spectra of the $^{85}$Rb-$^{87}$Rb mixture are 
shown in Fig.~\ref{f14}, in correspondence to the dynamics shown in Fig.~\ref{f08}. 
With the spectra calculated in the rotating frame, the upper row of panels refer to the spectra for 
the component $^{85}$Rb, with the lower row for the component $^{87}$Rb. The selected time
presented corresponds to the panels shown in the panels of Fig.~\ref{f08}.
As we have considered for the previous mixture, for this system we are also focusing in particular 
the region II, shown in Fig.~\ref{f11}, where the turbulent regime occurs. For that, we display the 
corresponding results in Fig.~\ref{f16}, by averaging over 50 samples in this region, considering 
fixed time intervals.
As comparing the two kind of mixtures, we observe that the Kolmogorov $k^{-5/3}$ power law 
behavior in the turbulent regime II (verified for $k\xi\lesssim 1$) is not essentially affected by the 
kind of mixtures we have. We also observe the same changing to $k^{-3}$ behavior for larger 
momenta.

\section{Conclusion}\label{sec5}
Vortex patterns produced dynamically by time-dependent elliptical laser stirring, together with the associated turbulent flow
behavior,  are studied and analyzed by considering two kinds of mass-imbalanced coupled Bose-Einstein condensates, within 
an effective two-dimensional pancake-like geometry.
For that, as illustrative coupled systems, we choose the $^{85}$Rb-$^{133}$Cs and $^{85}$Rb-$^{87}$Rb mixtures, which are 
understood as easily accessible and controllable systems in cold-atom experiments. 
The two-body inter- and intra-species
interactions are fixed along this study to be in a miscible configuration, $\delta=a_{12}/a_{11}=0.5$, with $a_{11}=a_{22}=60a_0$,
as considered more appropriate to investigate the time evolution of the coupled system in comparison with previous studies
with single component condensates.

Different regimes that occur in the time evolution due to laser stirring are analyzed, with the first stage being the shape 
deformation introduced by the stirring potential. This is followed by a symmetry breaking turbulent regime, in which 
vortex nucleation start to be generated at the surface of the two-species condensates.  A final regime in the time
evolution is verified with the production of stable vortex patterns,  associated to the assumed rotational frequency of 
the stirring potential.
Within an independent model simulation, these stirred-produced vortex patterns are than identified with similar 
patterns, which can be verified by direct ground-state calculations with classical rotational frequencies 
replacing the time-dependent stirring potential. For that, the classical rotational frequencies are derived from 
the expected values of the angular momentum and moment inertia operators, averaging in time.  

In order to quantify and characterize the vortex nucleations in the coupled condensates, when considering the stirring
periodic time-dependent potential, the time evolution of relevant dynamical observables are calculated for each one of the
two components of the mixture, such as the total and kinetic energies, the current densities and torques. 
The study of the kinetic energies is split in two parts: compressible, which is associated to acoustic wave generations;  
and incompressible one, which occurs due to vortex nucleation processes. Then, the increasing in the kinetic energies
is mainly associated to the compressible part in the turbulent interval till the time when vortices start to nucleate at
the surface, whereas along the following longer-time evolution, the energy increasing being predominantly due to the 
incompressible part. 
The dynamical process obtained with the stirring potential is concluded with the coupled system relaxing 
to the crystallization of lattice patterns, as one can also verify for single species condensates~\cite{2005Parker}. 
This similarity is more apparent in the coupled case $^{85}$Rb-$^{87}$Rb, when the mass difference is very small.
However, the mass-imbalanced effect can better be verified in the results obtained for the $^{85}$Rb-$^{133}$Cs mixture
(in which the higher-mass component has larger kinetic energy than the lower-mass one),
with the inter-species interaction being responsible for the production of a richer pattern of vortices, which are 
more visible for the higher-mass component. For comparison, see results for the densities in the panels
(d$_i$) of Figs.~\ref{f01} and ~\ref{f08}. 

In the dynamical evolution, it is noticed the initial torque experienced by the coupled condensates soon after 
switching on the laser stirring, being stronger for the less-massive element. With the vortices entering the system, 
going to stabilization of vortex patterns, the torques are saturated for both elements, being stronger for the 
more-massive one.
To calculate the effective time-dependent rotational frequency experienced by  both components due to laser stirring, 
we used the classical rotational relation, obtained from the expected values of the angular momenta and moment of inertia 
operators. 
By averaging the frequencies along the time evolution, the corresponding constant external rotations are defined 
approximately for the ground-state calculations, without the stirring potential.
The net related observation for coupled mixtures is that higher-mass condensed mixtures  
experience large rotational frequency than the lower-mass mixtures. 
This is also reflected in the visible number of vortices in the final patterns.

The sharp rising observed for the kinetic energy parts and time-dependent rotational frequencies, lead us to 
substantiate our investigation by considering the regimes due to laser stirring, within power spectra analysis in 
the momentum space, for the incompressible kinetic energy spectra in the rotating frame. 
As being verified, the incompressible velocity power spectra in the turbulent regime shows the $k^{-5/3}$ 
power-law behavior. In this regime, the vortex core satisfies the $k^{-3}$ power-law. 
The vortices in this stage are close to the boundary of the condensate. The $k^{-5/3}$ power-law vanishes 
after stable configurations of the vortices well entered into the condensate and the acoustic noises that 
affect the vortex core, where $k^{-3}$ power-law cannot be realized in the spectra. 
While comparing both mass-imbalanced mixtures, we noticed that higher mass imbalanced coupled condensates 
show the $k^{-5/3}$ power-law for a longer time window, than the near equal mass-imbalanced case. 
Regarding the mass-imbalanced comparative analysis, among the two kind of mass-imbalanced systems, 
we also  noticed that the change in the scaling behavior occurs at lower incompressible kinetic energies for larger mass differences between the species. Also, for such larger mass differences,  the dynamical production of stable patterns of 
vortices is verified to be a much faster process.

As a perspective, the present study for coupled condensates, under stirring potential, can be extended by using 
damped Gross-Pitaevskii analysis, with a treatment of the condensates with their corresponding thermal components.
It can be useful to calculate precisely the effective rotational velocity of the thermal clouds. 
 
 \section*{Acknowledgements}
ANS thanks partial support from 
Coordena\c c\~ao de Aperfei\c coamento de Pessoal de N\'\i vel Superior (CAPES).
RKK and ASB acknowledge support from the Marsden Fund of New Zealand [grant No. UOO1726].
LT acknowledges partial support from 
Funda\c{c}\~ao de Amparo \`a Pesquisa do Estado de S\~ao Paulo (FAPESP) 
[Contract 2017/05660-0] and Conselho Nacional de Desenvolvimento Cient\'\i fico e 
Tecnol\'ogico (CNPq) [Procs. 304469-2019-0].

\end{document}